\documentclass[envcountsame,runningheads]{llncs}
\usepackage{stmaryrd}
\usepackage{hyperref}
\usepackage{inconsolata}
\usepackage[T1]{fontenc}
\usepackage{soul}

\usepackage{paralist}
\usepackage{geometry}
\usepackage{bbm}
\usepackage{bbold}
\usepackage{listings}

\makeatletter
\newcommand{\xRightarrow}[2][]{\ext@arrow 0359\Rightarrowfill@{#1}{#2}}
\makeatother

\usepackage{varwidth}
\usepackage{enumitem}

\usepackage{amsmath}
\usepackage{amssymb} 
\usepackage{graphicx}
\usepackage{algorithm}
\usepackage{algorithmicx}
\usepackage[noend]{algpseudocode}
\usepackage{color}
\usepackage{xspace}
\usepackage{pgf}

\usepackage{tikz}
\usetikzlibrary{arrows,snakes,backgrounds,automata,shapes,positioning,chains,calc}

\usepackage[inference]{semantic}

\usepackage{marginfix} 

\newcommand{\OMIT}[1]{}

\usepackage{cutwin}
\usepackage{cite}
 
\newcommand{\porder}{\prec}
\newcommand{\dorder}{\prec\!\!\!\prec}
\newcommand{\zwr}[2]{\mathbb{0}_{#1}\,{#2}}
\newcommand{\amo}[2]{\mathbb{1}_{#1}\,{#2}}

\newcommand{\enc}[3]{#2 \stackrel{{\it enc}}{=}_{#1} #3}

\begin{document}


\title{Integrating Owicki-Gries for C11-Style Memory Models into
  Isabelle/HOL \thanks{This work is supported by EPSRC grants
    EP/R032351/1, EP/R032556/1 and EP/R019045/2.}}
%
%
\author{Sadegh Dalvandi\inst{1} \and  Brijesh
 Dongol\inst{1} \and Simon Doherty\inst{2}}
\institute{University of Surrey, Guildford, UK   \and University of Sheffield, Sheffield, UK}
%


\newcommand{\linefill}{\cleaders\hbox{$\smash{\mkern-2mu\mathord-\mkern-2mu}$}\hfill\vphantom{\lower1pt\hbox{$\rightarrow$}}}  
\newcommand{\Linefill}{\cleaders\hbox{$\smash{\mkern-2mu\mathord=\mkern-2mu}$}\hfill\vphantom{\hbox{$\Rightarrow$}}}  
\newcommand{\transi}[2]{\mathrel{\lower1pt\hbox{$\mathrel-_{\vphantom{#2}}\mkern-8mu\stackrel{#1}{\linefill_{\vphantom{#2}}}\mkern-11mu\rightarrow_{#2}$}}}
\newcommand{\trans}[1]{\transi{#1}{{}}}
\newcommand{\transo}{\mathord{\trans{~}}}
\newcommand{\ntransi}[2]{\mathrel{\lower1pt\hbox{$\mathrel-_{\vphantom{#2}}\mkern-8mu\stackrel{#1}{\linefill_{\vphantom{#2}}}\mkern-8mu\nrightarrow_{#2}$}}}
\newcommand{\ntrans}[1]{\ntransi{#1}{{}}}
\newcommand{\eseq}[1]{\langle~\rangle}

\newcommand{\traceArrow}[1]{\stackrel{#1}{\longrightarrow}}
\newcommand{\ltsArrow}[1]{\stackrel{#1}{\Longrightarrow}}

\newcounter{sarrow}
\newcommand\strans[1]{%
  \mathrel{\raisebox{0.1em}{
    \stepcounter{sarrow}%
    \!\!\!\!
    \begin{tikzpicture}
      \node[inner sep=.5ex] (\thesarrow) {$\scriptstyle #1$};
      \path[draw,<-,decorate,line width=0.25mm,
      decoration={zigzag,amplitude=0.7pt,segment length=1.2mm,pre=lineto,pre length=4pt}] 
      (\thesarrow.south east) -- (\thesarrow.south west);
    \end{tikzpicture}%
  }}}

\newcommand\Strans[1]{%
\mathrel{\raisebox{0.1em}{
\!\!\begin{tikzpicture}
  \node[inner sep=0.6ex] (a) {$\scriptstyle #1$};
  \path[line width=0.2mm, draw,implies-,double distance between line
  centers=1.5pt,decorate, 
    decoration={zigzag,amplitude=0.7pt,segment length=1.2mm,pre=lineto,
    pre   length=4pt}] 
    (a.south east) -- (a.south west);
\end{tikzpicture}}%
}}

\newcommand{\calE}{{\cal E}}
\newcommand{\nat}{\mathbb{N}}
\newcommand{\noteq}{\neq}

\newcommand{\lt}{{\bf Less than}}

\newcommand{\ev}{\mathit{ev}}
\newcommand{\Events}{\mathit{Evt}}
\newcommand{\Inv}{\mathit{Inv}}
\newcommand{\Resp}{\mathit{Res}}
\newcommand{\his}{\mathit{his}}
\newcommand{\exec}{\mathit{exec}}
\newcommand{\complete}{\mathit{complete}}
\newcommand{\Var}{\mathit{Var}}
\newcommand{\GVar}{{\it Var_G}}
\newcommand{\LVar}{\mathit{Var_L}}
\newcommand{\Val}{\mathit{Val}}
\newcommand{\Tid}{\mathit{Tid}}
\newcommand{\CASOp}{\mathit{CAS}} 
\newcommand{\WR}{\mathsf{W_R}}
\newcommand{\RA}{\mathsf{R_A}}
\newcommand{\R}{\mathsf{R}}
\newcommand{\A}{\mathsf{A}}
\newcommand{\RX}{\mathsf{R_X}}
\newcommand{\W}{\mathsf{W}}
\newcommand{\WX}{\mathsf{W_X}}
\newcommand{\C}{\mathsf{C}}
\newcommand{\CRA}{\mathsf{CRA}}
\newcommand{\URA}{\mathsf{U}}
\newcommand{\URAT}{\mathsf{UT}}
\newcommand{\URAF}{\mathsf{UF}}

\newcommand{\HB}{{\sf hb}\xspace} 
\newcommand{\PO}{{\sf po}\xspace}
\newcommand{\MO}{{\sf mo}\xspace}
\newcommand{\SC}{{\sf sc}\xspace}
\newcommand{\RF}{{\sf rf}\xspace}
\newcommand{\SB}{{\sf sb}\xspace}

\newcommand{\refeq}[1]{(\ref{#1})}
\newcommand{\refalg}[1]{Algorithm~\ref{#1}}

\newcommand{\fr}{{\sf fr}}
\newcommand{\ltsb}{{\sf sb}}
\newcommand{\ltrf}{\mathord{\sf rf}}
\newcommand{\ltfr}{{\sf fr}}
\newcommand{\lthb}{{\sf hb}}
\newcommand{\ltsw}{{\sf sw}}
\newcommand{\ltmox}{{\sf mo}^x}
\newcommand{\ltmo}{{\sf mo}}
\newcommand{\lteco}{{\sf eco}}
\newcommand{\PreExec}{{\it PreExec}}
\newcommand{\Approx}{{\it C11}}
\newcommand{\Seq}{{\it Seq}}

\newcommand{\True}{{\it true}}
\newcommand{\False}{{\it false}}

\newcommand{\justified}{justified\xspace}
\newcommand{\notjustified}{unjustified\xspace}
\newcommand{\Justified}{Justified\xspace}
\newcommand{\Notjustified}{Unjustified\xspace}

\newcommand{\rdval}{{\it rdval}}
\newcommand{\wrval}{{\it wrval}}
\newcommand{\loc}{{\it loc}}
\newcommand{\var}{\texttt{var}}

\newcommand{\imp}{\Rightarrow}
\newcommand{\expr}{\mathit{Exp}}
\newcommand{\flag}{\mathit{flag}}
\newcommand{\turn}{\mathit{turn}}
 
\newcommand{\hbo}[1]{\stackrel{#1}{\rightarrow}}
\newcommand{\detval}[1]{\stackrel{#1}{=}}
\newcommand{\last}{\mathit{last}}

\newcommand{\dview}{{\it dview}}
\newcommand{\wfs}{{\it wfs}}
\newcommand{\finite}{{\it finite}}

\newcommand{\kwswap}{\textsf{\textbf{swap}}}
\newcommand{\kwskip}{\textsf{\textbf{skip}}}
\newcommand{\kwdo}{\textsf{\textbf{do}}}
\newcommand{\kwwhile}{\textsf{\textbf{while}}}
\newcommand{\kwend}{\textsf{\textbf{end}}}
\newcommand{\kwif}{\textsf{\textbf{if}}}
\newcommand{\kwthen}{\textsf{\textbf{then}}}
\newcommand{\kwelse}{\textsf{\textbf{else}}}
\newcommand{\kwreturn}{\textsf{\textbf{return}}}
\newcommand{\kwthread}{\textsf{\textbf{thread}}}
\newcommand{\kwuntil}{\textsf{\textbf{until}}}

\newcommand{\whilestep}[1]{\stackrel{#1}{\longrightarrow}}
\newcommand{\fv}{\mathit{fv}}

\algnewcommand\Swap{\kwswap}
\algnewcommand\Skip{\kwskip}
\algnewcommand\Thread{\kwthread}

\algrenewcommand\algorithmicend{\kwend}
\algrenewcommand\algorithmicdo{\kwdo}
\algrenewcommand\algorithmicwhile{\kwwhile}
\algrenewcommand\algorithmicfor{\textsf{\textbf{for}}}
\algrenewcommand\algorithmicforall{\textsf{\textbf{for all}}}
\algrenewcommand\algorithmicloop{\textsf{\textbf{loop}}}
\algrenewcommand\algorithmicrepeat{\textsf{\textbf{repeat}}}
\algrenewcommand\algorithmicuntil{\textsf{\textbf{until}}}
\algrenewcommand\algorithmicprocedure{\textsf{\textbf{procedure}}}
\algrenewcommand\algorithmicfunction{\textsf{\textbf{function}}}
\algrenewcommand\algorithmicif{\kwif}
\algrenewcommand\algorithmicthen{\kwthen}
\algrenewcommand\algorithmicelse{\kwelse}
\algrenewcommand\algorithmicreturn{\kwreturn}

\algblockdefx{Thread}{EndThread}%
[1]{\kwthread \xspace #1}%
{\algorithmicend}

\algblockdefx{MyWhile}{EndMyWhile}%
[1]{\kwwhile \xspace #1}%
{\algorithmicend}

\makeatletter
\ifthenelse{\equal{\ALG@noend}{t}}%
  {\algtext*{EndMyWhile}}
  {}%
\makeatother

\algblockdefx{MyUntil}{EndMyUntil}%
[1]{\kwuntil \xspace #1}%
{\algorithmicend}
\makeatletter
\ifthenelse{\equal{\ALG@noend}{t}}%
  {\algtext*{EndMyUntil}}
  {}%
\makeatother

\makeatletter
\ifthenelse{\equal{\ALG@noend}{t}}%
  {\algtext*{EndThread}}
  {}%
\makeatother

\newcommand{\action}[3]{\ensuremath{
\begin{array}[t]{l@{~}l}
\multicolumn{2}{l}{#1}\\
\textsf{Pre:}&#2\\
\textsf{Eff:}&#3
\end{array}
}}

\newcommand{\kwtag}{{\it tag}}
\newcommand{\tid}{{\it tid}}
\newcommand{\act}{{\it act}}
\newcommand{\Op}{A}
\newcommand{\Prog}{{\it Prog}}
\newcommand{\Comm}{{\it Com}}
\newcommand{\AComm}{{\it ACom}}
\newcommand{\Exp}{{\it Exp}}
\newcommand{\Init}{\mathbf{Init}}

\newcommand{\Sur}{{\it C11 RAR}}
\newcommand{\asgn}{\ensuremath{:=}}

\newcommand{\bbD}{\mathbb{D}}
\newcommand{\bbE}{\mathbb{E}}
\newcommand{\bbS}{\mathbb{S}}

\newcommand{\rat}{\mathbb{Q}}
\newcommand{\bool}{\mathbb{B}}

\newcommand{\maxmo}{\mathbf{max}_{\ltmo}}
\newcommand{\cclose}{\mathbf{cclose}}
\newcommand{\scomp}{\circ}
\newcommand{\view}{\mathit{View}}
\newcommand{\tview}{\texttt{tview}}
\newcommand{\ls}{\mathit{ls}}
\newcommand{\rdview}{\mathit{rdview}}
\newcommand{\isReleasing}{\mathit{isReleasing}}
\newcommand{\mview}{\texttt{mview}}
\newcommand{\mods}{\texttt{mods}}
\newcommand{\writes}{\texttt{writes}}
\newcommand{\covered}{\texttt{covered}}
\newcommand{\val}{\texttt{val}}
\newcommand{\rel}{\texttt{rel}}

\newcommand{\tst}{{\tt tst}}
\newcommand{\OW}{\mathtt{OW}}
\newcommand{\visWrites}{\mathit{V\!W}\!}
\newcommand{\encounteredWrites}{\mathit{E\!W}\!}
\newcommand{\Act}{{\sf Act}}

\newcommand{\eqrng}[2]{(\ref{#1}-\ref{#2})}
\newcommand{\refprop}[1]{Proposition~\ref{#1}}
\newcommand{\reffig}[1]{Fig.~\ref{#1}}
\newcommand{\refthm}[1]{Theorem~\ref{#1}}
\newcommand{\reflem}[1]{Lem\-ma~\ref{#1}}
\newcommand{\refcor}[1]{Corollary~\ref{#1}}
\newcommand{\refsec}[1]{Section~\ref{#1}}
\newcommand{\refex}[1]{Example~\ref{#1}}
\newcommand{\refdef}[1]{Definition~\ref{#1}}
\newcommand{\reflst}[1]{Listing~\ref{#1}}
\newcommand{\refchap}[1]{Chapter~\ref{#1}}
\newcommand{\reftab}[1]{Table~\ref{#1}}

\newcommand{\WrX}[2]{#1 := #2}
\newcommand{\WrR}[2]{#1 :=^{\sf R} #2}
\newcommand{\RdX}[2]{#1 \gets #2}
\newcommand{\RdA}[2]{#1 \gets^{\sf A} #2}
\newcommand{\CObs}[5]{[#1 = #2](#4 =_{#3}{#5})}

\newcommand{\red}[1]{{\textcolor{red}{#1}}}

\newcommand\ogal{\{\!|}
\newcommand\ogar{|\!\}}

\newcommand\oga[1]{\ogal #1 \ogar}

\tikzset{
    mo/.style={dashed,->,>=stealth,thick,black!20!purple},
    hb/.style={solid,->,>=stealth,thick,blue},
    sw/.style={solid,->,>=stealth,thick,black!50!green},
    rf/.style={dashed,->,>=stealth,thick,black!50!green},
    fr/.style={dashed,->,>=stealth,thick,red}
    pa/.style={preaction={
         draw,yellow,-,
         double=yellow,
         double distance=1.5\pgflinewidth,
       }}    
 }

 \lstset{
    mathescape=true,
    breaklines=false,
    tabsize=2,
    morekeywords={if, then, else, let, in },keywordstyle=\color{blue},
    basicstyle=\ttfamily,
    literate={\ \ }{{\ }}1
  }

\maketitle

\begin{abstract} Weak memory presents a new challenge for program
  verification and has resulted in the development of a variety of
  specialised logics. 
  For C11-style memory models, our previous work has shown that it is
  possible to extend Hoare logic and Owicki-Gries reasoning to verify
  correctness of weak memory programs.  The technique introduces a set
  of high-level assertions over C11 states together with a set of
  basic Hoare-style axioms over atomic weak memory statements (e.g.,
  reads/writes), but retains all other standard proof obligations for
  compound statements. This paper takes this line of work further by
  showing Nipkow and Nieto's encoding of Owicki-Gries in the Isabelle
  theorem prover can be extended to handle C11-style weak memory
  models in a straightforward manner. 
  We exemplify our techniques over several litmus tests from the
  literature and a non-trivial example: Peterson's algorithm adapted
  for C11. For the examples we consider, the proof outlines can be
  automatically discharged using the existing Isabelle tactics
  developed by Nipkow and Nieto. The benefit here is that programs can
  be written using a familiar pseudocode syntax with assertions
  embedded directly into the program.
\end{abstract}



\section{Introduction}

Hoare logic~\cite{DBLP:journals/cacm/Hoare69} is fundamental to
understanding the intented design and semantics of sequential
programs. Owicki and Gries'~\cite{DBLP:journals/acta/OwickiG76}
framework extends Hoare logic to a concurrent setting by adding an
interference-free check that guarantees stability of assertions in one
thread against the execution of another. Although several other
techniques for reasoning about concurrent programs have since been
developed~\cite{DBLP:books/cu/RoeverBH2001}, Owicki-Gries reasoning
remains fundamental to understanding concurrent systems and one of the
main methods for performing deductive verification. Mechanised support
for Owicki-Gries' framework has been developed for the Isabelle
theorem prover~\cite{DBLP:books/sp/Paulson94} by Nipkow and
Nieto~\cite{DBLP:conf/fase/NipkowN99} and is currently included in the
standard distribution.

Our work is in the context of C11 (the 2011 C standard), which has a
weak memory model that is designed to enable programmers to take
advantage of weak memory
hardware~\cite{DBLP:conf/popl/BattyOSSW11,DBLP:conf/popl/KangHLVD17,DBLP:conf/pldi/LahavVKHD17,DBLP:journals/siglog/Lahav19}. Unlike
in sequentially consistent memory~\cite{DBLP:journals/tc/Lamport79},
states are graphs with several relations that are used to track
dependencies between memory events (e.g., reads, writes,
updates)~\cite{DBLP:conf/popl/AlglaveC17,DBLP:conf/popl/BattyOSSW11,DBLP:conf/ppopp/DohertyDWD19,DBLP:conf/popl/KangHLVD17,DBLP:conf/pldi/LahavVKHD17,DBLP:journals/siglog/Lahav19}. 
This means that it is not possible to use the traditional Owicki-Gries
framework to reason about concurrent programs under C11. Researchers
have instead developed a set of specialised logics, e.g.,
\cite{DBLP:conf/popl/AlglaveC17}, including those that extend
Owicki-Gries framework \cite{DBLP:conf/icalp/LahavV15} and separation
logic
\cite{DBLP:conf/vmcai/DokoV16,DBLP:conf/esop/SvendsenPDLV18,DBLP:conf/esop/SvendsenPDLV18,DBLP:conf/esop/DokoV17,DBLP:conf/oopsla/TuronVD14}
designed to cope with specific fragments of C11. 

Our point of departure is the operational semantics of Doherty et
al.~\cite{DBLP:conf/ppopp/DohertyDWD19} for the RC11-RAR fragment of
C11~\cite{DBLP:conf/pldi/LahavVKHD17}. As indicated by the \emph{RAR},
the memory model allows both relaxed and release-acquire
accesses. Moreover, the model restricts the C11 memory model to
disallow the ``load-buffering'' litmus
test~\cite{DBLP:conf/pldi/LahavVKHD17,DBLP:journals/siglog/Lahav19}.
A key advancement in the semantics developed by Doherty et al. is a
transition relation over states modelled as C11 graphs, allowing
program execution to be viewed as an interleaving of program
statements as in classical approaches to concurrency. 
They provide a primitive assertion language for expressing properties
of such states, which is manually applied to the message passing
litmus test and Peterson's algorithm adapted to C11. However, the
assertion language itself expresses state properties at a low level of
abstraction (high level of detail), and hence is difficult to
mechanise.  We\footnote{For the CAV reviewers: This work is under
  review. An anonymised copy of the paper has been
  uploaded~\cite{ESOP}.} have recently recast Doherty et al.'s
semantics in an equivalent timestamp-based
semantics\cite{DBLP:conf/popl/KangHLVD17,DBLP:conf/ecoop/KaiserDDLV17}. More
importantly, we have developed a high-level set of assertions for
stating properties of the C11 state~\cite{ESOP}. These assertions have
been shown to integrate well with a Hoare-style proof calculus, and,
by extension, the Owicki-Gries proof method. Interestingly, the
technique enables reuse of all standard Owicki-Gries proof rules for
compound statements.

In this paper, we push this technique further by showing how this
framework can be integrated into Isabelle via a straightforward
extension of the Owicki-Gries encoding by Nipkow and
Nieto~\cite{DBLP:conf/fase/NipkowN99}. Unlike~\cite{ESOP}, where
program counters are used to model control flow and relations over C11
states are used to model program transitions, the approach in this
paper is more direct. We show that once a correct proof outline has
been encoded, the proof outlines can be validated with minimal user
interaction. Our extension is parametric in the memory model, and can
be adapted to reason about other C11-style operational
models~\cite{DBLP:journals/siglog/Lahav19}.

\noindent
{\bf Contributions.} Our main contributions are thus: \vspace{-0.5em}
\begin{enumerate}
\item A simple and generic extension to the standard Isabelle encoding
  of Owicki-Gries to cope with C11-style weak memory,
\item An instantiation of the RC11-RAR operational semantics within
  Isabelle as an example memory model,
\item An integration with a high-level assertion language for
  reasoning about weak memory states, and
\item Verification of several examples in the extended theory,
  including Peterson's algorithm for C11.
\end{enumerate}\vspace{-0.5em}


\noindent {\bf Overview.} In \refsec{ogisabelle}, we briefly present
the Owicki-Gries encoding by Nipkow and
Nieto~\cite{DBLP:conf/fase/NipkowN99}, as well as the message passing
litmus test which serves as a running example. We describe how this
encoding can be generically extended to cope with weak memory in
\refsec{sec:lang_ext}, then in \refsec{sec:deduct-reas-weak}, we
present RC11-RAR as an example instantiation. In~\refsec{sec:verc11},
we present a technique for reasoning about C11-style programs as
encoded in Isabelle\footnote{Our entire Isabelle
  formalisation 
  can be found as ancillary files here
  \url{https://arxiv.org/abs/2004.02983}.}, which we apply to a number
  of examples. We present related work in \refsec{sec:related-work}.



\section{Owicki-Gries in Isabelle/HOL}
\label{ogisabelle}


Nipkow and Nieto~\cite{DBLP:conf/fase/NipkowN99} present a
formalisation of Owicki-Gries method in Isabelle/HOL. Their
formalisation defines syntax, its semantics and Owicki-Gries proof
rules in higher-order logic. Correctness of the proof rules with
respect to the semantics is proved and their formalisation is part of
the standard Isabelle/HOL libraries. To provide some context for our
extensions, we provide an overview of this encoding here; an
interested reader may wish to consult the original
paper~\cite{DBLP:conf/fase/NipkowN99} for further details.

The defined programming language is a C-like \texttt{WHILE} language
augmented with shared-variable parallelism ($||$) and synchronisation
(\texttt{AWAIT}). Parallelism must not be nested, i.e. within
$c_1 ~||~ c_2 ~|| ~...~ ||~ c_n$, each $c_i$ must be a sequential
program. The programming language allows program constructs to be
annotated with assertions in order to record proof outlines that can
later be checked. 
The language also allows unannotated commands that may be placed
within the body of {\tt AWAIT} statements. As in the original
treatment~\cite{DBLP:journals/acta/OwickiG76}, {\tt AWAIT} is an
atomic command that executes under the precondition of the {\tt AWAIT}
block.
\begin{small}
\begin{lstlisting}[escapeinside={(*}{*)}]
$\color{blue} \tt datatype$ (*$\alpha$*) ann_com =
AnnBasic ((*$\alpha$*) assn) ((*$\alpha$*) (*$\Rightarrow$*) (*$\alpha$*))
| AnnSeq ((*$\alpha$*) ann_com) ((*$\alpha$*) ann_com)
			("_ ;; _")
| AnnCond ((*$\alpha$*) assn) ((*$\alpha$*) bexp) ((*$\alpha$*) ann_com) ((*$\alpha$*) ann_com)
			("_ IF _ THEN _ ELSE _ FI")
| AnnWhile ((*$\alpha$*) assn) ((*$\alpha$*) bexp) ((*$\alpha$*) assn) ((*$\alpha$*) ann_com)
			("_ WHILE _ INV _ DO _ OD")
| AnnAwait ((*$\alpha$*) assn) ((*$\alpha$*) bexp) ((*$\alpha$*) com)
			("_ AWAIT _ THEN _ END")

$\color{blue} \tt datatype$ (*$\alpha$*) com =
Parallel ((*$\alpha$*) ann_com option (*$\times$*) (*$\alpha$*) assn) list
| Basic ((*$\alpha$*) (*$\Rightarrow$*) (*$\alpha$*))
| Seq ((*$\alpha$*) com)  ((*$\alpha$*) com) ("_ ,, _")
| Cond ((*$\alpha$*) bexp)  ((*$\alpha$*) com)  ((*$\alpha$*) com) ("IF _ THEN _ ELSE _ FI")
| While ((*$\alpha$*) bexp)  ((*$\alpha$*) assn)  ((*$\alpha$*) com) ("WHILE _ INV _ DO _ OD")
\end{lstlisting}
\end{small}

\noindent
In the datatype above, the concrete syntax is defined within 
\texttt{(" ... ")}. \texttt{$\alpha$ assn} and \texttt{$\alpha$ bexp}
represent assertions and Boolean expressions, respectively. 
\texttt{AnnBasic} represents a basic (atomic) state transformation
(e.g, an assignment). \texttt{AnnSeq} is sequential composition,
\texttt{AnnCond} is conditional, \texttt{AnnWhile} is a loop annotated
with an invariant, and \texttt{AnnWait} is a synchronisation
construct.  
The command \texttt{Parallel} is a list of pairs $(c, q)$ where $c$ is
an annotated (sequential) command and $q$ is a post-condition. The
concrete syntax for parallel composition (not shown above) is:
\texttt{COBEGIN $c_1~\oga{q_1} ~ || ~...~ ||~ c_n ~ \oga{q_n}$
  COEND}. 

The semantics of programs are defined by transition rules between
configurations, which are pairs comprising a program fragment and a
state.  
The proof rules of the Owicki-Gries formalisation are syntax directed.
A proof obligation generator has been implemented in the form of an
Isabelle tactic called {\tt oghoare}. Application of this tactic
results in generation of all standard Owicki-Gries proof obligations,
each of which can be discharged either automatically or via an
interactive proof. We omit the full details of standard semantics and
verification condition generator~\cite{DBLP:conf/fase/NipkowN99}.

\begin{example}[Message passing (MP)]\label{ref:MP}
  To motivate our extensions to a C11-style weak memory model, we
  consider the \emph{message passing} litmus test given in
  \reffig{fig:sc-mp}.  It comprises two shared variables: $d$ (that
  stores some data) and $f$ (that stores a flag), both of which are
  initially $0$.  Under sequential consistency, the postcondition of
  the program is $r2 = 5$. This is because the loop in thread~2 only
  terminates after $f$ has been updated to $1$ in thread 1, which in
  turn happens after $d$ has been set to $5$ by thread 1. Therefore,
  the only possible value of $d$ for thread 2 to read is $5$. The
  proof of this property straightforward, and can be easily handled by
  Nipkow and Nieto's encoding~\cite{DBLP:conf/fase/NipkowN99}. We
  provide a partial encoding in \reffig{fig:mp_og_sc} to provide an
  example instantiation of the abstract syntax above, and to better
  highlight the extensions necessary to handle C11-style weak memory.
  The state of the program is defined using an Isabelle record where
  all shared variables and local registers are modelled as its
  fields. For the proof outline in \reffig{fig:mp_og_sc}, the {\tt
    oghoare} tactic generates 29 proof obligations, each of which is
  automatically discharged.
\end{example}




\begin{figure}[t]
  \begin{minipage}[b]{0.45\columnwidth}
    \begin{center} 
      {\bf Init: } $d:=0;$ $f:=0;$  \qquad \qquad \qquad \qquad\\ 
      $\begin{array}{l@{\quad}||@{\quad }l}
         \begin{array}[t]{@{}l@{}}
           {\bf Thread }\ 1\\
           1: d := 5; \\
           2: f := 1;\\
         \end{array}
         & 
           \begin{array}[t]{@{}l@{}}
             {\bf Thread }\ 2\\
             3: \kwdo\ r1 := f \ \\
             \ \ \ \ \kwuntil\ r1 = 1;  \\ 
             4: r2 := d; \\
           \end{array}
       \end{array}$
       
       {\color{blue} $\{ r2=5\}$}  \qquad \qquad       
     \end{center}
     \caption[caption]{MP litmus test under sequential
       consistency 
     }
     \label{fig:sc-mp}
   \end{minipage}
   \hfill
   \begin{minipage}[b]{0.5\columnwidth}
     \begin{lstlisting}[escapeinside={(*}{*)}, lineskip = -5pt]
$\color{blue} \tt record$ mp_state =
d::nat   f::nat   r1::nat   r2::nat
$\color{blue} \tt lemma$:
||-  (*$\ogal$*) f = 0 (*$\land$*) d = 0 (*$ \ogar$*)
       COBEGIN
       (*$\ogal$*) True (*$ \ogar\ $*)  d := 5 ;;
       (*$\ogal$*) d = 5 (*$ \ogar$*) f := 1
       (*$\ogal$*) True (*$ \ogar$*)
       ||
       ...
       COEND
       (*$ \ogal$*) r2 = 5 (*$ \ogar$*) 
     \end{lstlisting}
     \vspace{-2em}
     \caption[caption]{Proving MP under sequential consistency using
       Nipkow and Nieto's encoding~\cite{DBLP:conf/fase/NipkowN99} of
       Owicki-Gries}
     \label{fig:mp_og_sc}
   \end{minipage}

 \end{figure}


\section{Extending Owicki-Gries to C11-Style Memory Models}
\label{sec:lang_ext}

We defer a precise description of a C11-style operational semantics to
\refsec{sec:deduct-reas-weak} in order to highlight the fact our
Isabelle framework is essentially parametric in the memory model used.
The fundamental requirement is that the memory model be an operational
model featuring C-style, annotated memory operations.  All that is
needed to understand the rest of this section is some basic
familiarity with weak memory
models~\cite{DBLP:conf/ppopp/DohertyDWD19,ESOP,DBLP:conf/ecoop/KaiserDDLV17,DBLP:journals/corr/PodkopaevSN16}. The
functions encoding the weak memory operational semantics {\tt WrX},
{\tt WrR}, {\tt RdX}, \dots will be instantiated
\refsec{sec:deduct-reas-weak}, and for the time being can be
considered to be transition functions that construct a new weak memory
state for a given weak memory prestate.  However, a reader may wish to
first read \refsec{sec:deduct-reas-weak} for an example C11 memory
model prior to continuing with the rest of this section.

To motivate our language extension, we reconsider MP (\refex{ref:MP})
in a C11-style weak memory model. In particular, if all reads and
writes are \emph{relaxed}, C11 admits an execution in which thread~2
reads a ``stale'' value of
$d$~\cite{DBLP:conf/ecoop/KaiserDDLV17,DBLP:conf/icalp/LahavV15}. Thus,
it is only possible to establish the weaker postcondition
$r2 = 0 \lor r2 = 5$ (see \refsec{sec:deduct-reas-weak} for details). To regain the expected
behaviour, one must introduce additional synchronisation in the
program. In particular, the write to $f$ in thread~1 must be a {\em
  releasing write} (denoted $\WrR{f}{1}$) and the read of $f$ in
thread~2 must be an {\em acquiring read} (denoted
$\RdA{r_1}{f}$).  

A weak memory state can be encoded as a special variable in the
standard semantics. Moreover, for the semantics that we
employ~\cite{DBLP:conf/ppopp/DohertyDWD19,ESOP}, within each weak
memory state, for each low-level weak memory event (e.g., read or
write) we must keep track of the thread identifier (of type {\tt T}),
the shared variable (or location) that is accessed (of type {\tt L})
and the value in that variable (of type {\tt V}).

The syntactic extensions necessary for encoding C11-style statements
in Isabelle are straightforward. For example, to capture the so-called
RAR fragment, we require five new programming constructs:
\emph{relaxed reads and writes}, \emph{releasing writes}, and
\emph{acquiring reads}. Moreover, we wish to support a {\tt SWAP[x,
  v]} command~\cite{PetersonBlog,DBLP:conf/ppopp/DohertyDWD19} that
acquiringly reads {\tt x} and releasingly writes {\tt v} to {\tt x} in
a single atomic step. This command is used in Peterson's algorithm
(see \reffig{fig:petersons_og_wm}) and is implemented in our model
using a \emph{read-modify-write update}.

All of the new extensions are defined using a shallow embedding and
their concrete syntax is enclosed in brackets \mbox{\texttt{< ... >}}
to avoid ambiguities in the Isabelle/HOL encoding. The annotated
versions of these statements are given below. For completeness, we
also require syntax for unannotated versions of each command, but
their details are elided.

\begin{lstlisting}
${\color{blue} \tt syntax}$
"_AnnWrX" :: "$\alpha$ assn $\Rightarrow$ L $\Rightarrow$ T  $\Rightarrow$ V $\Rightarrow$ Cstate $\Rightarrow$ $\alpha$ ann_com"  
  				("(_ <_  :=_  _ $\acute{}$_>)")
"_AnnWrR" :: "$\alpha$ assn $\Rightarrow$ L $\Rightarrow$ T  $\Rightarrow$ V $\Rightarrow$ Cstate $\Rightarrow$ $\alpha$ ann_com"   
  				("(_ <_ :=$\sf ^R$_  _ $\acute{}$_>)")
"_AnnRdX"::"$\alpha$ assn $\Rightarrow$ idt $\Rightarrow$ T $\Rightarrow$ L $\Rightarrow$ Cstate $\Rightarrow$ $\alpha$ ann_com"
                ("(_ < $\acute{}$_  $\leftarrow$_  _ $\acute{}$_ >)")  		
"_AnnRdA"::"$\alpha$ assn $\Rightarrow$ idt $\Rightarrow$ T $\Rightarrow$ L $\Rightarrow$ Cstate $\Rightarrow$ $\alpha$ ann_com"
                ("(_ < $\acute{}$_  $\leftarrow$$\sf ^A$_  _ $\acute{}$_ >)")
"_AnnSwap"::"$\alpha$ assn  $\Rightarrow$ L $\Rightarrow$ V $\Rightarrow$ T $\Rightarrow$ Cstate $\Rightarrow$ $\alpha$ ann_com"   
  				("(_ <SWAP[_, _]_ $\acute{}$_>)")
\end{lstlisting}

\noindent Antiquotations of the from
$\acute{}x$ are used to denote that the element
$x$ is a field of the record corresponding to the state of each
program. For example, for the program in \reffig{fig:mp_og_sc}, the
variables {\tt d}, {\tt f}, {\tt r1} and {\tt r2} of record {\tt
  mp\_state} would be interpreted in the syntax using antiquotated
variables. This allows the translations (see below) to update fields
in an external record yet for the programs themselves to be written
using a natural imperative syntax.

To cope with weak memory, we embed the weak memory state as a special
variable in the standard encoding (see Figs.~\ref{fig:lb_og_wm} and
\ref{fig:mp_og_wm}). Each operation induces an update to this embedded
weak memory state variable that can be observed by subsequent
operations on the weak memory state.

{\tt \_AnnWrX} defines a relaxed write. Its first argument is an
assertion (the precondition) of the command, the second is the
variable being modified, the third is the thread performing the
operation, the fourth is the value being written, and the fifth is the
weak memory prestate. Similarly, {\tt \_AnnWrA} is a releasing write.
{\tt \_AnnRdX} defines a relaxed read, which loads a value of the
given location (of type~{\tt L}) from the given weak memory prestate into
the second argument (of type {\tt idt}). An acquiring read, defined by
{\tt \_AnnRdA} is similar. Finally, {\tt \_AnnSwap} defines a swap
operation that writes the given value (third argument) to the given
location (second argument) using an update operation. 

The semantics of this extended syntax is given by a {\tt translation},
which updates the program variables, including the weak memory
state. For the commands above, after omitting some low-level Isabelle
details, we have:
\begin{lstlisting}
${\color{blue} \tt translations}$
"r < x :=$\tt \tt _t$ v $\acute{}$s>" $\rightharpoonup$
      "AnnBasic r (_update_name s (WrX $\acute{}$s x v t))"
"r < x :=$\sf ^R$$\tt _t$ v $\acute{}$s>" $\rightharpoonup$
      "AnnBasic r (_update_name s (WrR $\acute{}$s x v t))"
"r < $\acute{}$x  $\leftarrow$$\tt _t$ y $\acute{}$s>" $\rightharpoonup$ 
      "AnnAwait r ((_update_name s (fst (RdX $\acute{}$s y t))),,
                               (_update_name x (snd (RdX $\acute{}$s y t))))"		
"r < $\acute{}$x  $\leftarrow^{\sf A}$$\tt _t$ y $\acute{}$s>" $\rightharpoonup$ 
      "AnnAwait r ((_update_name s (fst (RdA $\acute{}$s y t))),,
                               (_update_name x (snd (RdA $\acute{}$s y t))))"		
"r <SWAP[x, v]$\tt _t$ $\acute{}$s>" $\rightharpoonup$ "AnnBasic r (fst (Upd $\acute{}$s x v t))"
\end{lstlisting}
These translations rely on an operational semantics defined by
functions {\tt WrX} (relaxed write), {\tt WrR} (releasing write), {\tt
  RdX} (relaxed read), {\tt RdA} (acquiring read) and {\tt Upd} (RMW
update), which we define in \refsec{sec:opsemisabelle}.

Relaxed and acquiring writes update the embedded weak memory state to
the state returned by {\tt WrX} and {\tt WrA}, respectively. A read
event must return a post state (which is used to update the value of
the embedded weak memory state) and the value read (which is used to
update the value of the local variable storing this value). In order
to preserve atomicity of the read, we wrap both updates within an
annotated {\tt AWAIT} statement. The translation of a {\tt SWAP} is
similar.

Note that a relaxed (acquiring) read comprises two calls to {\tt RdX}
({\tt RdA}), which one could mistakenly believe to cause two different
effects on the weak memory state. However, as we shall see, these
operations are implemented using Hilbert choice ({\tt SOME}), hence,
although there may be multiple values that a read could return, the
two applications of {\tt RdX} ({\tt RdA}) are identical for the same
value for the same parameters.

\section{A C11-Style Memory Model: RC11-RAR}
\label{sec:deduct-reas-weak}

\begin{figure}[t]
  \hfill
  \begin{minipage}[b]{0.48\columnwidth}
      \begin{center} 
  {\bf Init: } $d:=0;$ $f:=0;$  \qquad \qquad \qquad \qquad\\
  $\begin{array}{@{}l@{\quad}||@{\quad }l}
     {\bf Thread }\ 1
     & {\bf Thread }\ 2\\
     1: d := 5; \qquad & 
                       3:  \kwdo\ r1 \gets f \\
          
     2: f := 1; & \ \ \ \ \kwuntil\ r1 = 1;  \\ 
     & 4:  r2 \gets d; \\
     \end{array}$

   {\color{red} $\{r2 = 0 \lor r2=5\}$}  \qquad \quad  \quad     
 \end{center}\vspace{-1em}
 \caption{Unsynchronised MP under RC11-RAR}
 \label{fig:po-message-bad}

  \end{minipage}\hfill{}
  \hfill
  \begin{minipage}[b]{0.45\columnwidth}
  \begin{center} 
  {\bf Init: } $d:=0;$ $f:=0;$  \qquad \qquad \qquad \qquad\\ 
   $\begin{array}{l@{\quad}||@{\quad }l}
      \begin{array}[t]{@{}l@{}}
        {\bf Thread }\ 1\\
        5: d := 5; \\
        6: f :=^{\sf R} 1;\\
      \end{array}
      & 
      \begin{array}[t]{@{}l@{}}
        {\bf Thread }\ 2\\
        7: \kwdo\ r1 \gets^{\sf A} f \ \\
        \ \ \ \ \kwuntil\ r1 = 1;  \\ 
        8: r2 \gets d; \\
      \end{array}
     \end{array}$

   {\color{blue} $\{ r2=5\}$}  \qquad \qquad       
 \end{center}
 \vspace{-1em}
 \caption[caption]{MP with release-acquire synchronisation 
 }
 \label{fig:po-message}
\end{minipage}
\end{figure}

In this section, we describe a particular instance of a C11-style
memory model: the RC11-RAR fragment. This fragment disallows the load
buffering litmus
test~\cite{DBLP:conf/popl/BattyOSSW11,DBLP:journals/siglog/Lahav19,DBLP:conf/pldi/LahavVKHD17},
and all accesses are either relaxed, releasing or acquiring. It is
straightforward to extend the model to incorporate more sophisticated
notions such as release sequences and non-atomic accesses, but these
are not considered as the complications they induce detract from the
main contribution of our work. It is worth noting that RC11-RAR is
still a non-trivial fragment~\cite{DBLP:conf/ppopp/DohertyDWD19}.

\subsection{Message Passing}
\label{sec:message-passing}

To motivate the memory model, consider again the MP litmus test but
for RC11-RAR (Figs. \ref{fig:po-message-bad} and
\ref{fig:po-message}). In \reffig{fig:po-message-bad}, all accesses
are relaxed, and hence the program can only establish the weaker
postcondition $r2 = 0 \lor r2 = 5$ since it is possible for thread 2
to read $0$ for {\tt d} at line 4. In \reffig{fig:po-message}, the
release annotation (line 2) and the acquire annotation (line 3)
induces a \emph{happens-before} relation if the read of $f$ reads from
the write at line 2~\cite{DBLP:conf/popl/BattyOSSW11}. This in turn
ensures that thread 2 sees the most recent write to $d$ at line~5.

We use the operational semantics described in \cite{ESOP}, which
models the weak memory state using \emph{timestamped writes} and
\emph{thread view
  fronts}~\cite{DBLP:journals/corr/PodkopaevSN16,Dolan:2018:LDRF,DBLP:conf/ecoop/KaiserDDLV17,DBLP:conf/popl/KangHLVD17}. A
timestamp is a rational number that \emph{totally orders} the writes
to each variable. A viewfront records the \emph{timestamp} that a
thread has encountered for each variable --- the idea is that a thread
may read from \emph{any} write whose timestamp is no smaller than the
thread's current viewfront. Similarly, a write may be introduced at
any timestamp greater than the current viewfront. The only caveat when
introducing a write is that it may not be introduced directly after a
\emph{covered write} (see
\cite{DBLP:conf/ppopp/DohertyDWD19,ESOP}). This caveat is to ensure
atomicity of RMW operations. In particular, a write to a variable $x$
is covered whenever there is a RMW on $x$ that reads from the
write. In this instance, it would be unsound for another write to $x$
to be introduced between the write that is read and the RMW~(see
\cite{DBLP:conf/ppopp/DohertyDWD19} for further details).

\vspace{-0.5em}

\begin{figure}[t]
  \begin{minipage}[b]{0.24\columnwidth}
    \begin{center}
      \fbox{
  \begin{tikzpicture}[scale=0.65]
    \draw[thick,->] (0,4) -- (0,1); \draw[thick,->] (1,4) --
      (1,1);
    
      \coordinate (d0) at (0,3.5); \coordinate (d5) at (0,2);
      \coordinate (f0) at (1,3.5); \coordinate (f1) at (1,1.5);

      \coordinate (T1-1) at (2.2,4); \coordinate (T2-1) at (2.2,2.5);
    
      \draw (0,4) node[above] {$d$}; \draw (1,4) node[above] {$f$};
      \draw (T1-1) node[above] {T1 view}; \draw (T2-1) node[below] {T2
        view};

      \filldraw [black] (d0) circle (2pt) node[left, black]
      {$0$};
     
      \filldraw [black] (f0) circle (2pt) node[right, black,yshift=2mm]
      {$0$};

      \filldraw [black] (T1-1) circle (2pt);
    
      \filldraw [black] (T2-1) circle (2pt);

\path [draw, blue, thick] (d0) edge [bend right=20]
        (f0) (f0) edge [dashed] (T1-1);
 \path [draw, thick, red] (d0) -- (f0) (f0) edge
        [dashed] (T2-1);

      \end{tikzpicture}}

    $\sigma_0$
    \end{center}
    \end{minipage}       
  \begin{minipage}[b]{0.24\columnwidth}
    \begin{center}
      \fbox{
    \begin{tikzpicture}[scale=0.65]
      \draw[thick,->] (0,4) -- (0,1); \draw[thick,->] (1,4) --
      (1,1);
    
      \coordinate (d0) at (0,3.5); \coordinate (d5) at (0,2);
      \coordinate (f0) at (1,3.5); \coordinate (f1) at (1,1.5);

      \coordinate (T1-1) at (2.2,4); \coordinate (T2-1) at (2.2,2.5);
    
      \draw (0,4) node[above] {$d$}; \draw (1,4) node[above] {$f$};
      \draw (T1-1) node[above] {T1 view}; \draw (T2-1) node[below] {T2
        view};

      \filldraw [black] (d0) circle (2pt) node[left, black]
      {$0$};
     
      \filldraw [black] (f0) circle (2pt) node[right, black,yshift=2mm]
      {$0$};

      \filldraw [black] (d5) circle (2pt) node[left,
        black] {$5$};

      \filldraw [black] (T1-1) circle (2pt);
    
      \filldraw [black] (T2-1) circle (2pt);

      \path [draw, blue, thick] (d5) -- (f0) (f0) edge
        [dashed] (T1-1);
      \path [draw, thick, red] (d0) -- (f0) (f0) edge
        [dashed] (T2-1);

      \end{tikzpicture}}
    $\sigma_1$
    \end{center}
    \end{minipage}
      \begin{minipage}[b]{0.24\columnwidth}
        \begin{center}
                \fbox{
    \begin{tikzpicture}[scale=0.65]
      \draw[thick,->] (0,4) -- (0,1); \draw[thick,->] (1,4) --
      (1,1);
    
      \coordinate (d0) at (0,3.5); \coordinate (d5) at (0,2);
      \coordinate (f0) at (1,3.5); \coordinate (f1) at (1,1.5);

      \coordinate (T1-1) at (2.2,4); \coordinate (T2-1) at (2.2,2.5);
    
      \draw (0,4) node[above] {$d$}; \draw (1,4) node[above] {$f$};
      \draw (T1-1) node[above] {T1 view}; \draw (T2-1) node[below] {T2
        view};

      \filldraw [black] (d0) circle (2pt) node[left, black]
      {$0$};
     
      \filldraw [black] (f0) circle (2pt) node[right, black,yshift=2mm]
      {$0$};

      \filldraw [black] (d5) circle (2pt) node[left,
        black] {$5$}; 

       \filldraw [black] (f1) circle (2pt) node[right,
        black] {$1$}; 

      \filldraw [black] (T1-1) circle (2pt);
    
      \filldraw [black] (T2-1) circle (2pt);

       \path [draw, blue, thick] (d5) -- (f1) (f1) edge
         [dashed] (T1-1);
       \path [draw, thick, red] (d0) -- (f0) (f0) edge
         [dashed] (T2-1);

      \end{tikzpicture}}
    
    $\sigma_2$
      \end{center}
\end{minipage}
      \begin{minipage}[b]{0.24\columnwidth}
        \begin{center}
          \fbox{
    \begin{tikzpicture}[scale=0.65]
      \draw[thick,->] (0,4) -- (0,1); \draw[thick,->] (1,4) --
      (1,1);
    
      \coordinate (d0) at (0,3.5); \coordinate (d5) at (0,2);
      \coordinate (f0) at (1,3.5); \coordinate (f1) at (1,1.5);

      \coordinate (T1-1) at (2.2,4); \coordinate (T2-1) at (2.2,2.5);
    
      \draw (0,4) node[above] {$d$}; \draw (1,4) node[above] {$f$};
      \draw (T1-1) node[above] {T1 view}; \draw (T2-1) node[below] {T2
        view};

      \filldraw [black] (d0) circle (2pt) node[left, black]
      {$0$};
     
      \filldraw [black] (f0) circle (2pt) node[right, black,yshift=2mm]
      {$0$};

      \filldraw [black] (d5) circle (2pt) node[left,
        black] {$5$}; 

       \filldraw [black] (f1) circle (2pt) node[right,
        black] {$1$}; 

      \filldraw [black] (T1-1) circle (2pt);
    
      \filldraw [black] (T2-1) circle (2pt);

      \path [draw, blue, thick] (d5) -- (f1) (f1) edge
        [dashed] (T1-1);

\path [draw, thick, red] (d0) -- (f1) (f1) edge
       [dashed] (T2-1);
     \end{tikzpicture}}

       $\sigma_3$

      \end{center}
\end{minipage}
 \vspace{-1em}
 \caption{An execution of the unsynchronised message passing}
 \label{fig:po-message-bad-exec}

\end{figure}
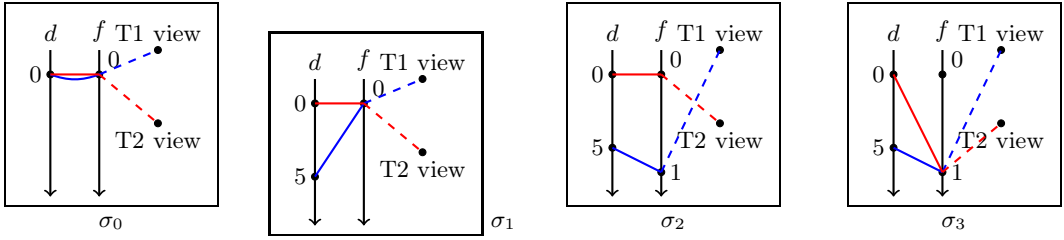

\begin{example}[Unsynchronised MP]
  \label{ex:unsync-mp}
  Consider \reffig{fig:po-message-bad-exec}, depicting a possible
  execution of the unsynchronised MP example
  (\reffig{fig:po-message-bad}). The execution comprises four weak
  memory states $\sigma_0$, $\sigma_1$, $\sigma_2$, $\sigma_3$. In
  each state, the timestamps themselves are omitted, but are assumed
  to be increasing in the direction of the arrows. The numbers depict
  the value of each variable at each timestamp. State $\sigma_0$ is
  the initial state. Each thread's viewfront in $\sigma_0$ is
  consistent with the initial writes.

  After executing line 1, the program transitions to $\sigma_1$, which
  introduces a new write (with value $5$) to $d$ and updates the
  viewfront of thread 1 to the timestamp of this write.  At this
  stage, thread 2's viewfront for $d$ is still at the write with value
  $0$. Thus, if thread 2 were to read from $d$, it would be permitted
  to return either $0$ or $5$. Moreover, if thread 2 were to write to
  $d$, it would be permitted to insert the write after $0$ or $5$.

  After executing line 2, the program transitions to $\sigma_2$, which
  installs a (relaxed) write of $f$ with value $1$. Now, consider the
  execution of line $3$. There are two possible poststates since there
  are two possible values of $f$ that thread 2 could read. State
  $\sigma_3$ depicts the case where thread 2 reads from the new write
  $f = 1$. In this case, the view front of thread 2 is updated, but
  crucially, since there is no release-acquire synchronisation, the
  viewfront of thread 2 for $d$ remains unchanged. This means that
  when thread 2 later reads from $d$ in line 4, it may return either
  $0$ or $5$. We contrast this with the execution of the synchronised
  MP described in \refex{ex:sync-mp}.
\end{example}

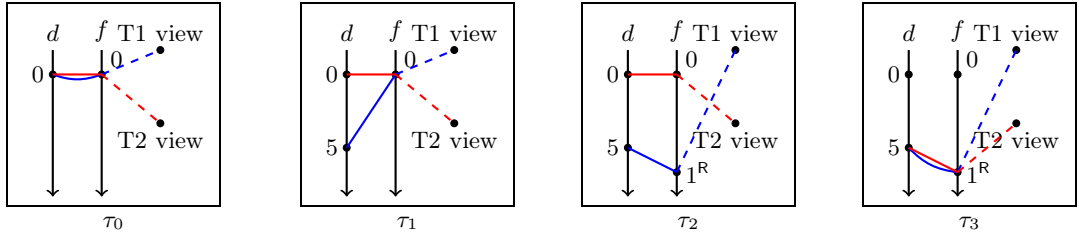
\begin{figure}[t]
  \begin{minipage}[b]{0.24\columnwidth}
    \begin{center}
      \fbox{
    \begin{tikzpicture}[scale=0.65]
      \draw[thick,->] (0,4) -- (0,1); \draw[thick,->] (1,4) --
      (1,1);
    
      \coordinate (d0) at (0,3.5); \coordinate (d5) at (0,2);
      \coordinate (f0) at (1,3.5); \coordinate (f1) at (1,1.5);

      \coordinate (T1-1) at (2.2,4); \coordinate (T2-1) at (2.2,2.5);
    
      \draw (0,4) node[above] {$d$}; \draw (1,4) node[above] {$f$};
      \draw (T1-1) node[above] {T1 view}; \draw (T2-1) node[below] {T2
        view};

      \filldraw [black] (d0) circle (2pt) node[left, black]
      {$0$};
     
      \filldraw [black] (f0) circle (2pt) node[right, black,yshift=2mm]
      {$0$};

      \filldraw [black] (T1-1) circle (2pt);
    
      \filldraw [black] (T2-1) circle (2pt);

\path [draw, blue, thick] (d0) edge [bend right=20]
        (f0) (f0) edge [dashed] (T1-1);
 \path [draw, thick, red] (d0) -- (f0) (f0) edge
        [dashed] (T2-1);

      \end{tikzpicture}}

    $\tau_0$
    \end{center}
    \end{minipage}       
    \hfill
  \begin{minipage}[b]{0.24\columnwidth}
    \begin{center}
      \fbox{
    \begin{tikzpicture}[scale=0.65]
      \draw[thick,->] (0,4) -- (0,1); \draw[thick,->] (1,4) --
      (1,1);
    
      \coordinate (d0) at (0,3.5); \coordinate (d5) at (0,2);
      \coordinate (f0) at (1,3.5); \coordinate (f1) at (1,1.5);

      \coordinate (T1-1) at (2.2,4); \coordinate (T2-1) at (2.2,2.5);
    
      \draw (0,4) node[above] {$d$}; \draw (1,4) node[above] {$f$};
      \draw (T1-1) node[above] {T1 view}; \draw (T2-1) node[below] {T2
        view};

      \filldraw [black] (d0) circle (2pt) node[left, black]
      {$0$};
     
      \filldraw [black] (f0) circle (2pt) node[right, black,yshift=2mm]
      {$0$};

      \filldraw [black] (d5) circle (2pt) node[left,
        black] {$5$};

      \filldraw [black] (T1-1) circle (2pt);
    
      \filldraw [black] (T2-1) circle (2pt);

      \path [draw, blue, thick] (d5) -- (f0) (f0) edge
        [dashed] (T1-1);
      \path [draw, thick, red] (d0) -- (f0) (f0) edge
        [dashed] (T2-1);

      \end{tikzpicture}}

    $\tau_1$
    \end{center}
    \end{minipage}
      \begin{minipage}[b]{0.24\columnwidth}
      \begin{center}
     \fbox{
    \begin{tikzpicture}[scale=0.65]
      \draw[thick,->] (0,4) -- (0,1); \draw[thick,->] (1,4) --
      (1,1);
    
      \coordinate (d0) at (0,3.5); \coordinate (d5) at (0,2);
      \coordinate (f0) at (1,3.5); \coordinate (f1) at (1,1.5);

      \coordinate (T1-1) at (2.2,4); \coordinate (T2-1) at (2.2,2.5);
    
      \draw (0,4) node[above] {$d$}; \draw (1,4) node[above] {$f$};
      \draw (T1-1) node[above] {T1 view}; \draw (T2-1) node[below] {T2
        view};

      \filldraw [black] (d0) circle (2pt) node[left, black]
      {$0$};
     
      \filldraw [black] (f0) circle (2pt) node[right, black,yshift=2mm]
      {$0$};

      \filldraw [black] (d5) circle (2pt) node[left,
        black] {$5$}; 

       \filldraw [black] (f1) circle (2pt) node[right,
        black] {$1^{\sf R}$}; 

      \filldraw [black] (T1-1) circle (2pt);
    
      \filldraw [black] (T2-1) circle (2pt);

      \path [draw, blue, thick] (d5) -- (f1) (f1) edge
        [dashed] (T1-1);
      \path [draw, thick, red] (d0) -- (f0) (f0) edge
        [dashed] (T2-1);
      \end{tikzpicture}}
    
       $\tau_2$
      \end{center}
\end{minipage}
      \begin{minipage}[b]{0.24\columnwidth}
        \begin{center}
          \fbox{
     
    \begin{tikzpicture}[scale=0.65]
      \draw[thick,->] (0,4) -- (0,1); \draw[thick,->] (1,4) --
      (1,1);
    
      \coordinate (d0) at (0,3.5); \coordinate (d5) at (0,2);
      \coordinate (f0) at (1,3.5); \coordinate (f1) at (1,1.5);

      \coordinate (T1-1) at (2.2,4); \coordinate (T2-1) at (2.2,2.5);
    
      \draw (0,4) node[above] {$d$}; \draw (1,4) node[above] {$f$};
      \draw (T1-1) node[above] {T1 view}; \draw (T2-1) node[below] {T2
        view};

      \filldraw [black] (d0) circle (2pt) node[left, black]
      {$0$};
     
      \filldraw [black] (f0) circle (2pt) node[right, black,yshift=2mm]
      {$0$};

      \filldraw [black] (d5) circle (2pt) node[left,
        black] {$5$}; 

       \filldraw [black] (f1) circle (2pt) node[right,
        black] {$1^{\sf R}$}; 

      \filldraw [black] (T1-1) circle (2pt);
    
      \filldraw [black] (T2-1) circle (2pt);

       \path [draw, blue, thick] (d5) edge [bend right=20]
        (f1) (f1) edge [dashed] (T1-1);

\path [draw, thick, red] (d5) -- (f1) (f1) edge         
       [dashed] (T2-1);

     \end{tikzpicture}}

   $\tau_3$
      \end{center}
\end{minipage}
 \vspace{-1em}
 \caption{An execution of the synchronised message passing}
 \label{fig:po-message-sync-exec}

\end{figure}

\begin{example}[Synchronised MP]\label{ex:sync-mp}
  Consider \reffig{fig:po-message-sync-exec}, which depicts an
  execution of the program in \reffig{fig:po-message}. State $\tau_1$
  is a result of executing line 5 and is identical to
  $\sigma_1$. However, after execution of line 6, we obtain state
  $\tau_1$, which installs a releasing write to $f$ (denoted by
  $1^{\sf R}$). As in \refex{ex:unsync-mp}, the acquiring read in
  line~7 could read from either of the writes to $f$. State $\tau_3$
  depicts the case in which thread 2 reads from the releasing write
  $1^{\sf R}$. Now, unlike \refex{ex:sync-mp}, this read establishes a
  release-acquire synchronisation, which means that the viewfront of
  thread 2 for \emph{both} $f$ and $d$ are updated. Thus, if the
  execution continues so that thread 2 reads from $d$ (line 8), the
  only possible value it may return is $5$.
\end{example}


\subsection{Operational Semantics of C11 RAR in Isabelle/HOL}
\label{sec:opsemisabelle}

We now present details of the memory model from
\refsec{sec:message-passing} as encoded in Isabelle/HOL. Recall that
the main purpose of this section is to instantiate the functions {\tt
  WrX}, {\tt WrR}, {\tt RdX}, {\tt RdA} and {\tt Upd} from
\refsec{sec:lang_ext}.

Recall that type \texttt{L} represents shared variables (or
locations), \texttt{T} represents threads, and \texttt{V} represents
values. We use type \texttt{TS} (which is synonymous with rational
numbers) to represent timestamps.  Each write can be uniquely
identified by a variable-timestamp pair. The type {\tt Cstate} is a
nested record with fields
\begin{itemize}
\item {\tt writes}, which is  the set of all writes,
\item {\tt covered}, which is the set of covered writes (recalling
  that covered writes are used to preserve atomicity of
  read-modify-write updates),
\item {\tt mods}, which is a function mapping each write to a write
  record (see below),
\item {\tt tview}, which is the viewfront (type {\tt
    L $\Rightarrow$ (L $\times$ TS)}) of each thread, and 
\item {\tt mview}, which is the viewfront of each write.
\end{itemize}
A write record contains fields {\tt val}, which is the value written
and {\tt rel}, which is a Boolean that is {\tt True} if, and only if,
the corresponding write is releasing.

\vspace{-1em}
\noindent
\begin{minipage}[t]{0.55\columnwidth}
\begin{lstlisting}
$\color{blue} \tt record$ Cstate = 
  writes::(L $\times$ TS) set
  covered::(L $\times$ TS) set
  mods::(L $\times$ TS) $\Rightarrow$ write_record
  tview::T $\Rightarrow$ L $\Rightarrow$ (L $\times$ TS)
  mview::(L $\times$ TS) $\Rightarrow$ L $\Rightarrow$ (L $\times$ TS)
\end{lstlisting}
\end{minipage}
\hfill
\begin{minipage}[t]{0.35\columnwidth}
\begin{lstlisting}
$\color{blue} \tt record$ write_record =
  val :: V
  rel :: bool	
\end{lstlisting}
\end{minipage}

\noindent 

Next, we describe how the operations modify the weak memory state.

\medskip\noindent{\textbf{Read transitions.}} Both relaxed and
acquiring reads leave all state components unchanged except for {\tt
  tview}. To define their behaviours, we first define a function {\tt
  visible\_writes $\sigma$ t x} that returns the set of writes to
$\tt x$ that thread $\tt t$ may read from in state $\sigma$. For a
write $\texttt{w} = \texttt{(x, q)}$, we assume a pair of functions
$\tt \var\ \texttt{w} = \texttt{x}$ and
$\tt \tst\ \texttt{w} = \texttt{q}$ that return the variable and
timestamp of $\texttt{w}$, respectively. Thus, we obtain:
\begin{lstlisting}
$\color{blue} \tt definition$ "visible_writes $\sigma$ t x $\equiv$
	 {w $\in$ writes $\sigma$ . var w = x $\wedge$ tst($\tview$ $\sigma$ t x) $\leq$ tst w}"
\end{lstlisting}
We use a function {\tt getVW} to select some visible write from which
to read:
\begin{lstlisting}
$\color{blue} \tt definition$ "getVW $\sigma$ t x $\equiv$
      (SOME w . w $\in$ visible_writes $\sigma$ t x)"
\end{lstlisting}
    
Finally, we require functions {\tt read\_transX t w $\sigma$} and {\tt
  read\_transA t w $\sigma$} that update the {\tt tview} component of
$\sigma$ for thread {\tt t} reading write {\tt w}.  Function {\tt
  read\_transX t w $\sigma$} updates $\tview\ \sigma\ \tt t$ to
$(\tview\ \sigma\ \tt t)[\var\ w := w]$, where $f [x := v]$ denotes
functional override. That is, the viewfront of thread {\tt t} for
variable {\tt var\ w} is updated to the write {\tt w} that {\tt t}
reads. The viewfronts of the other threads as well as the viewfront of
{\tt t} on variables different from {\tt var w} are unchanged. Thus,
the function {\tt RdX} required by the translation of a relaxed read
command in \refsec{sec:lang_ext} is thus defined by:
\begin{lstlisting}[mathescape=true,escapechar=+]
$\color{blue} \tt definition$ value $\sigma$ w $\equiv$ val (mods $\sigma$ w) +\vspace{-0.5\baselineskip}+
  
$\color{blue} \tt fun$ RdX :: "L  $\Rightarrow$ T  $\Rightarrow$ Cstate $\Rightarrow$ (Cstate $\times$ V)" where
       "RdX x t $\sigma$ =  (let w = getVW $\sigma$ t x; v = value $\sigma$ w in
                                  (read_transX t  w $\sigma$, v))"
\end{lstlisting}
We use {\tt value $\sigma$ w} to obtain the value of the write {\tt w}
in state $\sigma$.  The update defined by function
\texttt{read\_transA\ t w $\sigma$} for an acquiring read is
conditional on whether {\tt w} is a relaxed write. If {\tt w} is
relaxed, $\tview\ \sigma\ \tt t$ is updated to
$(\tview\ \sigma\ \tt t)[\var\ w := w]$ (i.e., behaves like a relaxed
read). Otherwise, the viewfront of {\tt t} must be updated to ``catch
up'' with the viewfront of {\tt w}. In particular,
$\tview\ \sigma\ \tt t$ is updated to
$(\tview\ \sigma\ \tt t) \otimes (\mview\ \sigma\ \tt w)$, where
\begin{align*}
  (f \otimes g)\ x  =
  \begin{cases}
    v_1\ x &\text{if}~\tst(v_2\ x) \leq \tst(v_1\ x)\\
    v_2\ x &\text{otherwise}
  \end{cases}
\end{align*}
Overall, we have:
\begin{lstlisting}
$\color{blue} \tt fun$ RdA :: "L  $\Rightarrow$ T  $\Rightarrow$ Cstate $\Rightarrow$ (Cstate $\times$ V)" where
       "RdA x t $\sigma$ =  (let w = getVW $\sigma$ t x; v = value $\sigma$ w in
                                  (read_transA t  w $\sigma$, v))"
\end{lstlisting}


\medskip\noindent{\textbf{Write transition.}} Writes update all state
components except {\tt covered}. First, following Doherty et
al.~\cite{DBLP:conf/ppopp/DohertyDWD19}, we must identify an existing
write {\tt w} in the current state; the new write is to be inserted
immediately after {\tt w}. Moreoever, {\tt w} must be visible to the
thread performing the write and covered by an RMW update. We define
the following function to
\begin{lstlisting}
$\color{blue} \tt definition$  "getVWNC $\sigma$ t x $\equiv$ 
	 SOME w . w $\in$ visible_writes $\sigma$ t x $\land$  w $\notin$ $\sigma$ covered"
\end{lstlisting}
where {\tt NC} stands for ``not covered''.  The write operation must
also determine a new timestamp, $\tt
ts$ for the new write. Given that the new write is to be inserted
immediately after the write operation {\tt w}, the timestamp $\tt
ts$ must be greater than $\tt tst\
w$ but smaller than the timestamp of other writes on $\var\ \tt
w$ after $\tt w$. Thus, we obtain a new timestamp using:
\begin{lstlisting}
$\color{blue} \tt definition$ "getTS $\sigma$ w $\equiv$ 
    SOME ts . tst w $<$ ts $\land$  
       ($\forall$ w' $\in$ writes $\sigma$. var w' = var w $\wedge$ tst w $<$ tst w' $\longrightarrow$
                                                       ts $<$ tst w')
\end{lstlisting}
Finding such a timestamp is always possible since timestamps are
rational numbers (i.e., are dense).

As with reads, we require a function {\tt write\_trans t b w v
  $\sigma$ ts} that updates the state $\sigma$ so that a new write
{\tt w' = ((var w), ts)} for thread {\tt t} is introduced with write
value {\tt v}. The Boolean {\tt b} is used to distinguish relaxed and
releasing writes. The write {\tt w} is the write after which the new
write {\tt w'} is to be introduced. The effect of {\tt write\_trans}
is to update $\writes\ \sigma$ to $\writes'$, $\mods\ \sigma$ to
$\mods'$ and both $\tview\ \sigma\ \tt t$ and $\mview\ \sigma\ \tt w'$
to $\tview'$, where:
\begin{align*}
  \writes' & =  (\writes\ \sigma) \cup \{\texttt{w'}\} \\
  \mods' & = (\mods\ \sigma\ \texttt{w'}) [\val := \texttt{v}, \rel := \texttt{b}] \\
  \tview' & = (\tview\ \sigma\ \texttt{t})[(\var\ {\tt w}) := \texttt{w'}] 
\end{align*}
Thus, $\writes'$ adds the new write $\texttt{w'}$ to the set of writes
of $\sigma$. The new $\mods'$ sets the value for $\texttt{w'}$ to {\tt
  v} and the $\rel$ field to {\tt b} (which is ${\tt True}$ iff the
new write $\texttt{w'}$ is releasing). Finally, $\tview'$ updates
$\tview$ of {\tt t} for variable $\var\ \texttt{w}$ (the variable that
both {\tt w} and {\tt w'} update) to $\texttt{w'}$.

Finally, the functions {\tt WrX} and {\tt WrR} required by the translations in \refsec{sec:lang_ext} are given as follows
\begin{lstlisting}
$\color{blue} \tt fun$ WrX :: "L $\Rightarrow$ V $\Rightarrow$ T $\Rightarrow$ Cstate $\Rightarrow$ Cstate" where
	"WrX x v t $\sigma$ = 
			(let w = getVWNC $\sigma$ t x ; ts' = getTS $\sigma$ w in 
								  write_trans t False w v $\sigma$ ts')"
$\color{blue} \tt fun$ WrR :: "L $\Rightarrow$ V $\Rightarrow$ T $\Rightarrow$ Cstate $\Rightarrow$ Cstate" where
	"WrR x v t $\sigma$ = 
			(let w = getVWNC $\sigma$ t x ; ts' = getTS $\sigma$ w in 
								  write_trans t True w v $\sigma$ ts')"
\end{lstlisting}

\noindent{\textbf{Update transition.}} Following Doherty et
al.~\cite{DBLP:conf/ppopp/DohertyDWD19}, we assume that an update
performs both an acquiring read and a releasing write in a single
step. It is possible to define variations that do not synchronise the
read or a write, but we omit such details for simplicity.

We first define a function \texttt{update\_trans t w v $\sigma$ ts}
that modifies $\sigma$ so that a releasing write {\tt w' = ((var w),
  ts)} by thread {\tt t} is introduced with write value {\tt v}
immediately after an existing write {\tt w}.  The effect of {\tt
  update\_trans} is to update $\writes\ \sigma$ to $\writes'$,
$\covered\ \sigma$ to $\covered'$, and $\mods\ \sigma$ to $\mods'$,
and both $\tview\ \sigma\ t$ and $\mview\ \sigma\ \texttt{w'}$ to
$\tview'$, where
\begin{align*}
  \writes' & =  (\writes\ \sigma) \cup \{\texttt{w'}\}
  \\
  \covered' & =  (\covered\ \sigma) \cup \{\texttt{w}\}
  \\
  \mods' & = (\mods\ \sigma \ \texttt{w'}) [\texttt{val} := \texttt{v}, \ \texttt{rel} := \texttt{True}]
  \\
  \tview' & = 
      \begin{cases}
        (\tview\ \sigma\ t)[(\var\ \texttt{w}) := \texttt{w'}] \otimes (\mview\ \sigma\ 
        \texttt{w}) 
        &\mbox{if ${\tt rel}\ (\mods\ \sigma\ \texttt{w})$ 
         }\\
         (\tview\ \sigma\ t)[(\var\ \texttt{w}) := \texttt{w'}]&\mbox{otherwise}
      \end{cases} 
\end{align*}
Thus, $\writes'$ adds the new write \texttt{w'} corresponding to the
update to the set of writes of $\sigma$ and $\covered'$ adds the write
{\tt w} that {\tt w'} reads from to the covered writes set of
$\sigma$. The new $\mods'$ sets the value for $\texttt{w'}$ to {\tt v} and
sets the {\tt rel} field to {\tt True}. Finally, $\tview'$ updates
$\tview$ of {\tt t} in the same way as a read operation, except that the
first case is taken provided the write {\tt w} being read is releasing.






The function {\tt Upd} required by the translation in
\refsec{sec:lang_ext} is given as follows:
\begin{lstlisting}
$\color{blue} \tt fun$  Upd :: "L $\Rightarrow$ V $\Rightarrow$ T $\Rightarrow$ Cstate $\Rightarrow$ (Cstate $\times$ V)" where
"Upd x v t $\sigma$ =  (let w = getVWNC $\sigma$ t x ; ts = getTS $\sigma$ w ;
                                       v = value $\sigma$ w in
                                              (update_trans t w v $\sigma$ ts, v))"
\end{lstlisting}

\noindent{\textbf{Well formedness.}} \refsec{sec:verc11} presents an
assertion language for verifying C11 programs. The lemmas introduced
there require states to be \emph{well formed}, which we characterise
by predicate \texttt{wfs} defined below.
\begin{lstlisting}[escapechar=+]
$\color{blue} \tt definition$  "writes_on $\sigma$ x $\equiv$ {w. var w = x $\wedge$ w $\in$ writes $\sigma$}" +\vspace{-0.5\baselineskip}+

$\color{blue} \tt definition$  "lastWr $\sigma$ x $\equiv$ (x, Max (tst`(writes_on $\sigma$ x)))"+\vspace{-0.5\baselineskip}+

$\color{blue} \tt definition$ "wfs $\sigma$ $\equiv$
                  ($\forall$ t x. $\tview$ $\sigma$ t x $\in$ writes_on $\sigma$ x) $\land$
                  ($\forall$ w x. mview $\sigma$ w x $\in$ writes_on $\sigma$ x) $\land$
                  ($\forall$ x. finite(writes_on $\sigma$ x)) $\land$
                  ($\forall$ w. w $\in$ writes $\sigma$ $\longrightarrow$ mview $\sigma$ w (var w) = w) $\land$
                  ($\forall$ w x. w = lastWr $\sigma$ x $\longrightarrow$ w $\notin$ covered $\sigma$)"

\end{lstlisting}
Function {\tt writes\_on $\sigma$ x} returns the set of writes in
$\sigma$ to variable {\tt x}.  Function {\tt lastWr $\sigma$ x}
returns the write on {\tt x} whose timestamp is greater than the
timestamp of all other writes on {\tt x} in state $\sigma$. 

In the definition of \texttt{wfs} $\sigma$, the first two conjuncts
ensure that all writes recorded in $\tview$ and $\mview$ are
consistent with \texttt{writes} $\sigma$. The third ensures the set of
writes in $\sigma$ is finite and the fourth ensures that for each
write in $\sigma$, the write's modification view of the variable it
writes is the write itself. The final conjuct ensures that the last
write to each variable (i.e., the write with the largest timestamp) is
not covered. We have shown that \texttt{wfs} is stable under each of
the transitions {\tt WrX}, {\tt WrR}, \dots. Thus the well-formedness
assumption made by the lemmas in \refsec{sec:verc11} is trivially
guaranteed.





\newcommand{\llp}{\llparenthesis}
\newcommand{\rrp}{\rrparenthesis}
\newcommand{\asigma}{\acute{}\sigma}
\newcommand{\ar}{$\acute{}$r}

\section{An Assertion Language for Verifying C11 Programs}
\label{sec:verc11}

In the previous sections, we discussed how the existing Owicki-Gries
theories in Isabelle can be extended with a weak memory C11
operational semantics in order to reason about C11-style programs
using standard proof rules. We mentioned that how a novel set of
assertions introduced in \cite{ESOP} can be used in our extension to
annotate programs w.r.t. C11 state and reason about them. In this
section we introduce the assertion language and present their
encodings in Isabelle through a number of examples and litmus
tests. We also provide some of the rules (lemmas) that Isabelle uses
to discharge proof obligations and validate the proof outlines. We
show how C11 state is incorporated into the programs and shared
variables are defined.  We also present a fully verified encoding of
the Peterson's mutual exclusion algorithm to further validate our
approach.


\subsection{Load Buffering}

\begin{figure}[t]

\begin{minipage}[b]{0.46\columnwidth}
  \small
\begin{lstlisting}[escapeinside={(*}{*)}, lineskip = -2pt,xleftmargin=2em, numbers=left, numberstyle=\tiny]
$\color{blue} \tt consts$
	x :: L   y :: L

$\color{blue} \tt record$ lb_state =
	r1 :: V
	r2 :: V
	(*$\sigma$*) :: Cstate

||-  
(*\color{red}{$\ogal$ \ar1 = 0 $\land$ \ar2 = 0 $\land$}*)
  (*\color{red}{  [x =$_1$ 0] $\asigma$ $\land$ [x =$_2$ 0] $\asigma$ $\land$ }*)
  (*\color{red}{  [y =$_1$ 0] $\asigma$ $\land$ [y =$_2$ 0] $\asigma$ $\ogar$*)
COBEGIN
	(*\color{red}{$\ogal$ [y =$_2$ 0] $\asigma$ $\land$  \ar2  = 0 $\ogar$*)
	< $\acute{}$r1 $\leftarrow$$_1$ x $\asigma$ > ;;
	(*\color{red}{$\ogal$ [y =$_2$ 0] $\asigma$ $\land$  \ar2  =  0 $\ogar$*)
	< y :=$_1$ 1 $\asigma$ >
	(*\color{red}{$\ogal$ \ar1 = 0 $\lor$ \ar2 = 0 $\ogar$*)
||
	(*\color{red}{$\ogal$ [x =$_1$ 0] $\asigma$ $\land$  \ar1  = 0 $\ogar$*)
	< $\acute{}$r2 $\leftarrow$$_2$ y $\asigma$ > ;;
	(*\color{red}{$\ogal$ [x =$_1$ 0] $\asigma$ $\land$  \ar1  = 0 $\ogar$*)
	< x :=$_2$ 1 $\asigma$ >
	(*\color{red}{$\ogal$ \ar1 = 0 $\lor$ \ar2 = 0 $\ogar$*)
COEND
(*\color{red}{$\ogal$ \ar1 = 0 $\lor$ \ar2 = 0 $\ogar$*)
\end{lstlisting}\vspace{-1em}
\caption[caption]{Isabelle encoding of the load-buffering litmus
  test.}
 \label{fig:lb_og_wm}
\end{minipage}
\hfill 
\begin{minipage}[b]{0.49\columnwidth}
\small 
\begin{lstlisting}[escapeinside={(*}{*)}, lineskip = -5pt,xleftmargin=2em, numbers=left, numberstyle=\tiny]
$\color{blue} \tt consts$ 
 d :: L     f :: L

$\color{blue} \tt record$ mp_state =
 r1 :: V    r2 :: V    $\sigma$::Cstate
	
||-   
(*\color{red}{$\ogal$ [f =$_1$ 0] $\asigma$ $\land$ [f =$_2$ 0] $\asigma$ $\land$*)
   (*\color{red}{[d =$_1$ 0] $\asigma$ $\land$ [d =$_2$ 0] $\asigma$ $\ogar$*)
COBEGIN
    (*\color{red}{$\ogal$ $\neg$[f $\approx$$_2$ 1] $\asigma$ $\land$ [d =$_1$ 0] $\asigma$ $\ogar$*)
	<d :=$_1$ 5 $\asigma$> ;;
    (*\color{red}{$\ogal$ $\neg$[f $\approx$$_2$ 1] $\asigma$ $\land$ [d =$_1$ 5] $\asigma$ $\ogar$*)
	<f :=$\sf ^R$$_1$ 1 $\asigma$>
    (*\color{red}{$\ogal$ True $\ogar$*)
||
	DO (*\color{red}{$\ogal$ [f = 1]$\llp$d =$_2$ 5$\rrp$ $\asigma$ $\ogar$  *)
		  < $\acute{}$r1 $\leftarrow$$\sf ^A$$_2$ f $\asigma$>
	UNTIL ( $\acute{}$r1 = 1)
	INV (*\color{red}{$\ogal$ [f = 1]$\llp$d =$_2$ 5$\rrp$ $\asigma$ $\land$ *)
	        (*\color{red}{( $\acute{}$r1 = 1 $\longrightarrow$ [d =$_2$ 5] $\asigma$) $\ogar$*)
	OD;;
    (*\color{red}{$\ogal$ [d =$_2$ 5] $\asigma$$\ogar$ *)
    < $\acute{}$r2 $\leftarrow$$_2$ d $\asigma$>
    (*\color{red}{$\ogal$ \ar2 = 5 $\ogar$*)
COEND 
(*\color{red}{$\ogal$ \ar2 = 5 $\ogar$*)
\end{lstlisting}\vspace{-1em}
 \caption[caption]{Isabelle encoding of the message-passing
  litmus test.}
 \label{fig:mp_og_wm}
\end{minipage}
\end{figure}

Our first example is the load-buffering litmus test given in
\reffig{fig:lb_og_wm}. As discussed earlier, we use restricted C11
memory model described by Lahav et
al.~\cite{DBLP:conf/pldi/LahavVKHD17}, and hence we prevent the
program from terminating by reading 1 for both {\tt x} in thread 1 and
{\tt y} in thread 2.

As mentioned earlier, the C11 state is represented as a field of the
record corresponding to the state of the program (i.e. as a field of
{\tt lb\_state} record for the load-buffering example). Updates to
$\sigma$ are via the underlying definition of the operations in
accordance with the RC11-RAR operational semantics as described in
\refsec{sec:opsemisabelle}. In our encoding, shared variables are
represented as constants representing locations in the C11 state
($\sigma$). 



Now consider the proof outline. The first assertion (lines 10-12)
specifies the initial state of the program. The first two conjuncts
are assertions on the value of local registers. The other four
conjuncts are \emph{definite observation} assertions. A definite
observation assertion denoted {\tt [x =\textsubscript{t} n] $\sigma$}
states that thread {\tt t}'s viewfront is consistent with the last
write to {\tt x} in $\sigma$ and that this write has value {\tt
  n}. Thus, if {\tt t} reads from {\tt x}, it is guaranteed to return
{\tt n}.
$\sigma$. 
Formally,
\begin{lstlisting}[escapeinside={*}{*}]
 [x =*\textsubscript{t}* n] $\sigma$ $\equiv$
		 tview $\sigma$ t = lastWr $\sigma$ x $\land$ value $\sigma$ (lastWr $\sigma$ x) = n
\end{lstlisting}
All weak-memory assertions in the proof outline
of~\reffig{fig:lb_og_wm} are definite value assertions, and this is
sufficient to prove the postcondition. However, to discharge the
generated proof obligations, we require the following two proof rules
over C11 assertions, which are defined as Isabelle lemmas:
\vspace{-1em}

\noindent
\begin{small}
\begin{minipage}[t]{0.44\columnwidth}
\begin{lstlisting}[escapechar=§]
$\color{blue} \tt lemma$ d_obs_RdX_pres:  
assumes "wfs $\sigma$"
and "[x =$_{\tt t}$ u] $\sigma$"
shows "[x =$_{\tt t}$ u]
             (fst(RdX y t' $\sigma$))"
\end{lstlisting}
\end{minipage}
\hfill 
\small
\begin{minipage}[t]{0.55\columnwidth}
\begin{lstlisting}[escapechar=§]
$\color{blue} \tt lemma$ d_obs_WrX_diff_var_pres:  
  assumes "wfs $\sigma$"
      and "[x =$_{\tt t}$ u] $\sigma$"
      and "y $\neq$ x"
      shows"[x =$_{\tt t}$ u] (WrX y v t' $\sigma$)"
\end{lstlisting}
\end{minipage}
\end{small}
\noindent Lemmas {\tt d\_obs\_RdX\_pres} and {\tt
  d\_obs\_WrX\_diff\_var\_pres} give conditions for pre\-serving
definite value assertions for relaxed read and write transitions,
respectively for an arbitrary variable {\tt y} and thread {\tt
  t'}. Note {\tt d\_obs\_WrX\_diff\_var\_pres} requires that the
variable {\tt y} that is written to is different from the variable
{\tt x} that appears in the definite value assertion. Both lemmas are
proved sound with respect to the operational semantics in
\refsec{sec:opsemisabelle}. Of course {\tt d\_obs\_RdX\_pres} also
holds for an acquiring read transition and {\tt
  d\_obs\_WrX\_diff\_var\_pres} for a releasing write
transition\footnote{In fact, our Isabelle theory provides a generic
  lemma that applies to both cases simultaneously.}.

The assertions on lines 14 and 20 are locally correct because of the
initial state. The assertions on lines 16 and 22 are locally correct
using {\tt d\_obs\_RdX\_pres}. Local correctness of the assertions on
lines 18 and 24 is trivial follows by the definite value
assertion. Interference freedom of the assertions in lines 14, 16, 20
and 22 is also established using the two lemmas.


\subsection{Message-Passing}

We now consider the assertions used in the message-passing algorithm
(\reffig{fig:mp_og_wm}). The first new assertion used in the proof
outline is the \textit{possible observation} assertion. This assertion
(denoted {\tt [x $\approx$\textsubscript{t} n] $\sigma$}) states that
thread {\tt t} may read value {\tt n} if it reads from variable {\tt
  x}. The formal definition in Isabelle is as follows:
\begin{lstlisting}[escapeinside={*}{*}]
[x $\approx$*\textsubscript{t}* n] $\sigma$ $\equiv$ $\exists$ w.w $\in$ visible_writes $\sigma$ t x $\land$ n = value $\sigma$ w
\end{lstlisting}
The next assertion we introduce is the \textit{conditional
  observation} assertion (denoted {\tt [x = n]$\llp$y
  =\textsubscript{t} m$\rrp$ $\sigma$}) which states that if thread
{\tt t} reads a value {\tt n} using an acquiring read for {\tt x}, it
synchronises with the corresponding write and obtains the definite
value {\tt m} for {\tt y}. Note that this requires that any write to
{\tt x} with value {\tt n} that {\tt t} can read is a releasing
write. The formal definition in Isabelle is as follows:
\begin{lstlisting}[escapeinside={*}{*}]
[x = n]$\llp$y =$_{\tt t}$ m$\rrp$ $\sigma$ $\equiv$ 
   $\forall$ w $\in$ visible_writes $\sigma$ t x. value $\sigma$ w = n $\longrightarrow$ 
		 mview $\sigma$ w y = lastWr $\sigma$ y  $\land$ value $\sigma$ (lastWr $\sigma$ y) = m 
               $\land$ rel (mods $\sigma$ w)"
\end{lstlisting}

Here we only introduce two of the
interesting rules used in the proof, and refer the interested reader
to the Isabelle theories for the remaining lemmas:\vspace{-1em}

\noindent 
\begin{small}
\begin{minipage}[t]{0.45\columnwidth}
\begin{lstlisting}[escapechar=§]
$\color{blue} \tt lemma$ c_obs_intro:  
assumes "wfs $\sigma$"
and "[y =$_{\tt t}$ m] $\sigma$"
and "$\neg$[x $\approx$$_{\tt t'}$ n] $\sigma$"
and "x $\neq$ y"
and "t $\neq$ t'"
shows "[x = u]$\llp$y =$_{\tt t'}$ v$\rrp$
                 (WrR x u t $\sigma$)"
\end{lstlisting}
\end{minipage}
\hfill 
\begin{minipage}[t]{0.53\columnwidth}
\begin{lstlisting}[escapechar=§]
$\color{blue} \tt lemma$ c_obs_transfer:
assumes "wfs $\sigma$"
and "[x = u]$\llp$y =$_{\tt t}$ v$\rrp$ $\sigma$"
and "snd(RdA $\sigma$ x t) = u"
shows "[y =$_{\tt t}$ v] (fst(RdA x t $\sigma$))"
\end{lstlisting}
\end{minipage}
\end{small}




\noindent 
Consider the conditional observation assertion in line 17. Local
correctness holds trivially by initialisation. Interference freedom
under line 12 is straightforward. For interference freedom under line
14, we use {\tt c\_obs\_intro}. In particular, the assertion at line
13 (i.e., the precondition of line 14) satisfies the critical premises
of {\tt c\_obs\_intro}. We use the conditional observation assertion
(line~17 of thread~2) in combination with rule {\tt c\_obs\_transfer}
to establish a definite observation on a new variable in thread~2. We
note that the variable read by the transition in rule {\tt
  c\_obs\_transfer} is {\tt x}, whereas the definite value assertion
in the consequent is on variable {\tt y}. Full details of this proof
may be found in \cite{ESOP}; in this paper, we focus on automation,
which we discuss in \refsec{subsect:verif-strategy}.

\subsection{Read-Read Coherence}
We have verified three versions of the read-read coherence (RRC)
litmus test using our extended theories. We have provided the more
interesting of the three in \reffig{fig:rrc3}. The other two versions
are provided Appendix \ref{sec:appendix1}. In order to prove this
example, a richer set of assertions is required. In particular, in
addition to the assertions regarding observability of writes, we need
assertions about the order of writes and the limits on the occurrence
of values.

The first assertion used for this example that we discuss here is the
\textit{possible value order} assertion (denoted \texttt{[m}
$\porder_\texttt{x}$ \texttt{n]} $\sigma$), which states that there
exists a write to variable {\tt x} with value {\tt n} ordered after
(i.e., with a greater timestamp) a write to {\tt x} with value {\tt
  m}. Similarly, a \textit{definite value order} assertion (denoted
\mbox{{\tt [m $\dorder_\texttt{x}$ n]} $\sigma$}) states that all
writes to {\tt x} with value {\tt n} are ordered after all writes to
{\tt x} with value {\tt m}. These are formally defined in Isabelle as
follows.
\begin{lstlisting}[escapechar=+]
[m $\porder_\texttt{x}$ n] $\sigma$ $\equiv$
  $\exists$ w w'.  w $\in$ writes_on $\sigma$ x $\land$  w' $\in$ writes_on $\sigma$ x $\land$
		  value $\sigma$ w = m $\land$ value $\sigma$ w' = n $\land$ (tst w) < (tst w')+\vspace{-0.5\baselineskip}+

[m $\dorder_\texttt{x}$ n] $\sigma$ $\equiv$ 
      ($\forall$ w w'.  w $\in$ writes_on $\sigma$ x  $\land$  w' $\in$ writes_on $\sigma$ x $\land$ 
        value $\sigma$ w = m $\land$ value $\sigma$ w' = n $\longrightarrow$ (tst w) < (tst w')) 
	$\land$ [m $\porder_\texttt{x}$ n] $\sigma$
\end{lstlisting}

The other two assertions appear in this proof outline fall into the
\textit{value occurrence} category:
{\tt [}$\zwr{\texttt{x}}{\texttt{n}}${\tt]}$_\texttt{i}$ means that there has not
been a write with value {\tt n} to variable {\tt x} (where {\tt i} is
the initial value of {\tt x}) and {\tt[}$\amo{\texttt{x}}{\texttt{n}}${\tt]}
means that there has been at most one write with value {\tt n} to {\tt
  x}. These two assertions are defined in terms of ordering assertions
introduced earlier as follows. The predicate {\tt init $\sigma$ x n}
holds iff the initial value of \texttt{x} in $\sigma$ is \texttt{n}.

\begin{lstlisting}[escapechar=+]
$\color{blue} \tt definition$ init $\sigma$ x n $\equiv$ 
  $\exists$ w . w $\in$ writes_on $\sigma$ x $\wedge$  value $\sigma$ w = n $\wedge$
      ($\forall$ w' $\in$ writes_on $\sigma$ x .  w $\neq$ w' $\longrightarrow$   (tst w) < (tst w')) +\vspace{-0.5\baselineskip}+

[$\zwr{\texttt{x}}{\texttt{n}}$]$_\texttt{i}$ $\sigma$ $\equiv$ init $\sigma$ x n $\wedge$ $\neg$[i $\porder_\texttt{x}$ n] +\vspace{-0.5\baselineskip}+

[$\amo{\texttt{x}}{\texttt{n}}$] $\sigma$ $\equiv$ $\neg$[n $\porder_\texttt{x}$ n]
\end{lstlisting}
  
The last new assertion used in this proof outline is
\textit{encountered value} (denoted as {\tt
  [}$\enc{\texttt{t}}{\texttt{x}}{\texttt{n}}${\tt ]}) means that
thread {\tt t} has had the opportunity to observe a write with value
{\tt n} of {\tt x}. This assertion is formally defined in Isabelle as
follows\footnote{Note that some of the notation in our Isabelle
  encoding is different. We use the notation from~\cite{ESOP} here for
  a cleaner presentation.}:

\begin{lstlisting}[escapeinside={*}{*}]
 [$\enc{\texttt{t}}{\texttt{x}}{\texttt{n}}$] $\sigma$ $\equiv$  $\exists$ w . w $\in$ writes_on $\sigma$ x $\land$ 
                             tst(w) $\leq$ tst(tview $\sigma$ t x) $\land$ value $\sigma$ w = n
\end{lstlisting}

The five assertions above, together with other assertions introduced
earlier, are sufficient to specify the behaviour of the three-threaded
version of RRC. The conditional observation assertion on line 18, is
used to capture the possible synchronisation between threads 1 and
2. The ordering assertions in thread 2 and 3 specify that if the
writes have happened in a specific order, the read order must remain
coherent with respect to the order of writes. Namely, if thread 2
synchronises with thread 1 (i.e., {\tt r1} is set to 1), then it must
have observed the write of {\tt x} at line 12. Thus, the write to {\tt
  x} with value 2 at line 23 must have happened after. Therefore, it
must be impossible for the third thread to read value 2 for {\tt x} at
line 27, then subsequently read 1 for {\tt x} at line 29. This
reasoning is captured by the postcondition of the program.


\begin{figure}[t]

\begin{minipage}[b]{\columnwidth}
  \small
\begin{lstlisting}[escapeinside={*}{*}, multicols=2,lineskip = -5pt,xleftmargin=2em, numbers=left, numberstyle=\tiny]
$\color{blue} \tt consts$
	x :: L      y :: L	
$\color{blue} \tt record$ rrc3_state =
	a :: V   b :: V    r1 :: V
	$\sigma$ :: Cstate

||-
*\color{red}{$\ogal$[x $=_2 0$] $\asigma$ $\land$ [y $=_3 0$] $\asigma$$\ogar$}*
COBEGIN
*\color{red}{$\ogal$[$\zwr{\texttt{x}}{1}$]$_0$ $\asigma$ $\land$ $\neg$[y $\approx$$_2$ 1] $\asigma$ $\land$ }*
    *\color{red}{([$\zwr{\texttt{x}}{2}$]$_0$ $\asigma$ $\longrightarrow$ [x =$_1$ 0] $\asigma$)$\ogar$}*
	*<x :=$_1$ 1  $\asigma$> ;;*
*\color{red}{$\ogal$$\neg$ [y $\approx$$_2$ 1] $\asigma$ $\land$ }*
    *\color{red}{([$\zwr{\texttt{x}}{2}$]$_0$ $\asigma$ $\longrightarrow$ [x =$_1$ 1] $\asigma$)$\ogar$}*
	*<y :=$\sf ^R$$_1$ 1  $\asigma$>*
*\color{red}{$\ogal$ True $\ogar$}*
||
*\color{red}{$\ogal$[$\zwr{\texttt{x}}{2}$]$_0$ $\asigma$ $\land$ [y = 1]$\llp$x =$_2$ 1$\rrp$ $\asigma$$\ogar$}*
	*< $\acute{}$r1 $\leftarrow$$\sf ^A$$_2$ y  $\asigma$> ;;*
*\color{red}{$\ogal$( $\acute{}$r1 = 1 $\longrightarrow$ [ x =$_2$ 1] $\asigma$ $\land$ }*
                        *\color{red}{[$\enc{2}{\texttt{x}}{1}$] $\asigma$) $\land$ }*
    *\color{red}{[$\zwr{\texttt{x}}{2}$]$_0$ $\asigma$$\ogar$}*
	*<x :=$_2$ 2  $\asigma$>*
*\color{red}{$\ogal$ $\acute{}$r1 = 1 $\longrightarrow$ [1 $\dorder_\texttt{x}$ 2] $\asigma$$\ogar$}*
||
*\color{red}{$\ogal$ True $\ogar$}*
	*< $\acute{}$a $\leftarrow_3$ x  $\asigma$> ;;*
*\color{red}{$\ogal$ [$\enc{3}{\texttt{x}}{\acute{}\texttt{a}}$]  $\asigma$ $\ogar$}*
	*< $\acute{\;}$b $\leftarrow_3$ x  $\asigma$>*
*\color{red}{$\ogal$( $\acute{}$a $\neq$ $\acute{\;}$b $\longrightarrow$ [ $\acute{}$a $\porder_\texttt{x}$ $\acute{\;}$b] $\asigma$)$\ogar$}*
COEND
*\color{red}{$\ogal$ $\acute{}$r1 = 1 $\land$ $\acute{}$a = 2 $\longrightarrow$ $\acute{\;}$b $\neq$ 1$\ogar$}*

\end{lstlisting}\vspace{-1em}
\caption[caption]{Isabelle encoding of the read-read coherence litmus
  test with three threads. The proof additionally relies on a global
  invariant {\tt [init x 0] $\asigma$ $\wedge$ [init y 0] $\asigma$
    $\wedge$ [$\amo{\texttt{x}}{1}$] $\asigma$ $\wedge$
    [$\amo{\texttt{x}}{2}$] $\asigma$}}
 \label{fig:rrc3}
\end{minipage}
\end{figure}

\subsection{Peterson's Algorithm for C11}
\begin{figure}[t]
\begin{minipage}[p]{\textwidth}
\small 
\begin{lstlisting}[escapeinside={*}{*}, lineskip = -10pt ]
||- 
*\color{red}{$\ogal$ $\acute{}$after1 $\land$  $\neg$ $\acute{}$after2 $\land$ [flag1 =$_1$ 0] $\asigma$ $\land$ [flag2 =$_2$ 0] $\asigma$  $\land$}*
    *\color{red}{[turn =$_1$ 0] $\asigma$ $\land$ [turn =$_2$ 0] $\asigma$ $\ogar$*
COBEGIN
  *\color{red}{$\ogal$ $\acute{}$$\neg$ after1 $\land$ [flag1 =$_1$ 0] $\asigma$ $\land$ (cvd[turn, 0] $\asigma$  $\lor$ cvd[turn, 1] $\asigma$) $\land$}*
     *\color{red}{($\acute{}$after2 $\longrightarrow$ cvd[turn, 1] $\asigma$ $\land$ [turn = 1]$\llp$flag2 =$_1$ 1$\rrp$ $\asigma$) $\land$ }*
     *\color{red}{$\neg$ [turn $\approx$$_2$ 2] $\asigma$  $\ogar$}*

  <flag1 :=$_1$ 1 $\asigma$>;;

  *\color{red}{$\ogal$ $\neg$ $\acute{}$after1 $\land$ [flag1 =$_1$ 1] $\asigma$ $\land$ $\neg$ [turn $\approx$$_2$ 2] $\asigma$   $\land$}*
     *\color{red}{($\neg$ $\acute{}$after2 $\longrightarrow$ cvd[turn, 1] $\asigma$ $\land$ [turn = 1]$\llp$flag2 =$_1$ 1$\rrp$ $\asigma$) $\ogar$}*

  $\langle$<SWAP[turn, 2]$_1$ $\asigma$>,, $\acute{}$after1 := True$\rangle$ ;;

  DO 
    *\color{red}{$\ogal$ $\acute{}$after1 $\land$ ($\acute{}$after2 $\land$ ([flag2 $\approx$$_1$ 0]  $\asigma$   $\lor$ [turn $\approx$$_1$ 1]  $\asigma$) $\longrightarrow$  [turn =$_2$ 1]  $\asigma$)$\ogar$}*

   < $\acute{}$r1 $\leftarrow$$\sf ^A$$_1$ flag2 $\asigma$>;;

    *\color{red}{$\ogal$ $\acute{}$after1 $\land$ }*
        *\color{red}{($\acute{}$after2 $\land$ ( $\acute{}$r1 = 0 $\lor$ [turn $\approx$$_1$ 1] $\asigma$ $\lor$ [flag2 $\approx$$_1$ 0] $\asigma$) $\longrightarrow$}*
                                                                                                     *\color{red}{[turn =$_2$ 1] $\asigma$)$\ogar$}*

    < $\acute{}$r2 $\leftarrow$$_1$ turn $\asigma$>  

  UNTIL ( $\acute{}$r1 = 0 $\lor$ $\acute{}$r2 = 1)
  INV
  *\color{red}{$\ogal$ $\acute{}$after1   $\land$ }*
      *\color{red}{($\acute{}$after2 $\land$ ( $\acute{}$r1 = 0 $\lor$ $\acute{}$r2 = 1 $\lor$ [turn $\approx$$_1$ 1] $\asigma$  $\lor$ [flag2 $\approx$$_1$ 0] $\asigma$) $\longrightarrow$}*
                                                                                                      *\color{red}{[turn =$_2$ 1] $\asigma$) $\ogar$}*
  OD ;;
  *\color{red}{$\ogal$ $\acute{}$after1  $\land$ ($\acute{}$after2  $\longrightarrow$ [turn =$_2$ 1] $\asigma$)$\ogar$}*

  $\langle$<flag1 :=$\sf ^R$$_1$ 0 $\asigma$>,, $\acute{}$after1 := False$\rangle$

  *\color{red}{$\ogal$ True $\ogar$}*
  
||  Thread 2 (symmetric)
COEND
*\color{red}{$\ogal$ True $\ogar$}*
\end{lstlisting}\vspace{-1em}
\caption[caption]{Proof outline for Peterson's
  algorithm under C11. The second thread (not shown here) is
  symmetric.}
 \label{fig:petersons_og_wm}
\end{minipage}
\end{figure}

The final non-trivial example in this paper is Peterson's mutual
exclusion algorithm (\reffig{fig:petersons_og_wm}) as described
in~\cite{PetersonBlog}.  The proof outline for this algorithm has the
new assertion {\em covered} (denoted {\tt cvd[x, n] $\sigma$}). The
assertion {\tt cvd[x, n] $\sigma$} means that every write to {\tt x}
in state $\sigma$ except the last is covered and the value written by
that last write is {\tt n}. This assertion is formally defined in
Isabelle as:
\begin{lstlisting}[escapeinside={*}{*}, lineskip = -5pt]
cvd[x, n] $\sigma$ $\equiv$
	$\forall$ w. w $\in$ writes $\sigma$  $\land$ var $\sigma$ w = x $\land$ w $\not\in$ covered $\sigma$ $\longrightarrow$ 
													w = lastWr $\sigma$ x $\land$ value $\sigma$ w = n
\end{lstlisting}
\noindent Similar to the previous examples, in order to prove the
Peterson's algorithm we will need to introduce new proof rules to deal
with assertions involving {\tt cvd}.

The proof also uses \emph{auxiliary variables} {\tt after1} and {\tt
  after2}. We set {\tt after1} to {\tt True} when {\tt turn} is set in
thread 1 and to {\tt False} when thread 1 exits its critical section
(i.e., when the flag is reset). This is a standard technique used in
Owicki-Gries proofs of Peterson's algorithm in the conventional,
sequentially consistent,
setting~\cite{DBLP:series/txcs/AptBO09,DBLP:conf/fase/NipkowN99}. Note
that introduction of auxiliary variables must follow the same rules as
the classical setting~\cite{DBLP:journals/acta/OwickiG76} and must not
be a shared constant that appears within the weak memory state
$\sigma$. This avoids the notions of unsoundness of auxiliary
variables described in earlier
work~\cite{DBLP:conf/icalp/LahavV15}. 

\subsection{Verification Strategy}
\label{subsect:verif-strategy}

For each of the algorithms described above, we employ a generic
verification strategy and outline the the proof effort.
After encoding the algorithm and the assertions, the main steps in
each validating the proof outlines are as follows.
\begin{enumerate}
\item First, we use the built-in {\tt oghoare} tactic to reduce an
  Owicki-Gries proof outline into a set of basic Hoare logic proof
  obligations over weak memory pre-post state assertions. This tactic
  is exactly as developed by Nipkow and Nieto
  \cite{DBLP:conf/fase/NipkowN99}, and is used without change.

\item 
  We pipe this output (which is a set of proof obligations on atomic
  commands) to the Isabelle simplifier, transforming the set-based
  representation of assertions by Nipkow and Nieto into a
  predicate-based representation. This allows the lemmas for weak
  memory that we haved adapted from \cite{ESOP} to be automatically
  matched with the proof obligations.
\end{enumerate}
For the simple litmus tests, the first two steps either discharge all
the proof obligations, or leave a few (maximum 6) proof obligations
unproved. These proof obligations require slightly more sophisticated
application of the lemmas over weak memory state than can be
discharged by the simplifier alone. However, they can be automatically
discharged using Isabelle's built-in {\tt sledgehammer} tool
\cite{DBLP:conf/cade/BohmeN10}. 

This verification strategy works equally well for Peterson's algorithm
although it is a larger example that generates a significantly higher
number of proof obligations. The {\tt oghoare} tactic generates 258
subgoals, but over half of these are discharged by step 2 above,
leaving 91 subgoals. Although automatic, repeated applications of {\tt
  sledgehammer} to discharge so many proof obligations is rather
tedious. However, one can quickly discover common patterns in the
proof steps allowing these proof obligations to be discharged via a
few simple applications of {\tt apply}-style proofs.

\section{Related Work}
\label{sec:related-work}
As has been mentioned, the current
paper builds on ideas found in \cite{DBLP:conf/ppopp/DohertyDWD19}. That paper
did not develop a program logic based on Hoare triples, and was limited to
invariance style proofs. Both \cite{DBLP:conf/ppopp/DohertyDWD19} and
the current paper use the same definite value assertion, but the current paper
employs a much richer and more powerful assertion language. In particular,
the conditional value assertion is critical for enabling an Owicki-Gries based
program logic. Finally, \cite{DBLP:conf/ppopp/DohertyDWD19} does not consider
mechanisation or automation.

Of course, a great deal of work has been put into the development of separation
logics for C11-style weak memory models
\cite{DBLP:conf/oopsla/TuronVD14, vafeiadis2013relaxed, DBLP:conf/vmcai/DokoV16, doko2017tackling,
DBLP:conf/ecoop/KaiserDDLV17}.
One of the mosty recent and perhaps most fundamental of these is the Iris
framework \cite{DBLP:conf/ecoop/KaiserDDLV17}. This framework has been formalised in
the Coq proof assistant, and instantiations of it support a large fragemt of C11.
This fragment contains C11's {\em nonatomic accesses} but not relaxed accesses and is
therefore incomparable to our own. In particular, nonatomic access cannot legally race,
whereas relaxed accesses are designed to enable racy code.
More generally, separation logics can
become complicated when applied to weak-memory,
and we are partly motivated by the desire to build verification frameworks atop
simple and natural relational models (other authors \cite{DBLP:conf/icalp/LahavV15} have
made similar observations).

There have been a number of recent attempts to develop mechanised deductive
verification support for weak memory. Summers and M{\"u}ller
\cite{DBLP:conf/tacas/Summers018} present an approach to automating
deductive verification for weak memory programs by encoding Relaxed Separation Logic
\cite{vafeiadis2013relaxed} and Fenced Separation Logic
\cite{DBLP:conf/vmcai/DokoV16, doko2017tackling} 
into Viper \cite{muller2016viper}. Their work builds on separation logic, whereas
ours builds on a relational framework.
Apart from this fundamental difference, Summers and M{\"u}ller encode the concurrent
logics into the Viper sequential specification framework, which provides a high level
of automation. On the other hand, and as the authors themselves note, encoding the
logic in a foundational verification tool such as Isabelle provides a higher
level of assurance about correctness. In particular, the entire development
of our framework is verified in Isabelle, down to the operational semantics.

Another technique based on Owicki-Gries is that of \cite{DBLP:conf/icalp/LahavV15},
which defines a proof system for the {\em release-acquire} fragment of C11,
a smaller fragment than the {\em release-acquire-relaxed} fragment that we treat in this
paper. It is unclear how difficult it would be to extend \cite{DBLP:conf/icalp/LahavV15}
to deal with relaxed accesses. In any case, \cite{DBLP:conf/icalp/LahavV15} does not deal
with mechanisation or automation.

Alglave and Cousot have developed another extension to the Owicki-Gries
method for weak memory models \cite{DBLP:conf/popl/AlglaveC17}.
Because their method, like ours, is an extension of the Owicki-Gries method,
their verification method first requires a proof outline.
One novelty of their approach is that their method requires
a {\em communication specification} (or CS), which involves specifying for each
read operation in the program, which writes the variable can read from
(which may be in another thread). Verifying that the proof outline and CS are together
valid, and therefore that assertions of the proof outline do in fact hold for the program,
involves two proof obligations. The first is to show that the proof outline is valid
in our standard sense (so that it is locally correct and noninterfering), under the assumption
that the CS is satisfied. The second obligation
is that a given memory model satisfies the CS. This second obligation constitutes
an additional proof effort, not required in our method. The advantage they gain
is that once a proof outline is known to be locally correct and noninterfering under a given CS,
then the algorithm is known to be correct under any memory model that satisfies the CS.

The operational semantics in the current paper is inspired by the
semantics described in
\cite{DBLP:conf/ecoop/KaiserDDLV17,DBLP:journals/corr/PodkopaevSN16}. The
current paper is based on semantics and assertions found in
\cite{ESOP}, which also presents case-study verifications mechanised
in Isabelle. The mechanisation in that paper is rudimentary. 
Programs are not represented in a while-style
language as in the current paper. Instead, they use a program-counter
based representation, where control flow must be explicitly modelled.
As a consequence, proof obligations are not decomposed in the conventional
Owicki-Gries style. Rather, the verifier must prove stability of a large global invariant
mapping program counter locations to the assertions that hold at that location.
Furthermore, there is little real automation, either in generating or
discharging proof obligations. The current paper presents a
highly-structured and mechanised Owicki-Gries framework supporting a
high level of automation.


Dan et al. \cite{dan2017effective} introduce an abstraction for the store
buffers of the weak memory model which reduces the workload on program
analysers. They provide a source-to-source transformation that
realises the abstraction producing a program that can be analysed with
verifiers for sequential consistency. The approach is integrated with
\textsc{ConcurInterproc} \cite{jeannet2013relational} and uses the Z3
theorem prover.  {\em Model checking} has also been targeted for weak
memory, e.g., by explicitly encoding architectural structures leading
to weak behaviour, like store
buffers~\cite{DBLP:conf/hvc/TravkinMW13,DBLP:conf/esop/AlglaveKNT13}.
Ponce de Le\'{o}n et
al.~\cite{DBLP:conf/fmcad/LeonFHM18,DBLP:conf/cav/GavrilenkoLFHM19}
have developed a bounded model checker for weak memory models, taking
the axiomatic description of a memory model as input.  (Bounded) model
checkers for specific weak memory models are furthermore the tools
CBMC~\cite{DBLP:conf/cav/AlglaveKT13} (for TSO), {\sc
  Nidhugg}~\cite{DBLP:journals/acta/AbdullaAAJLS17} (for TSO and PSO),
RCMC~\cite{DBLP:journals/pacmpl/Kokologiannakis18} (for C11) and {\sc
  Gen}MC~\cite{DBLP:conf/pldi/Kokologiannakis19}.
Others \cite{dalvandi2019towards} present an approach for modelling and
verifying C11 programs using Event-B and ProB model checker.

\section{Conclusion}
\label{sec:conclusion}

In this paper, we have introduced an extension to a twenty-year old
formalisation of Owicki-Gries proof calculus in
Isabelle~\cite{DBLP:conf/fase/NipkowN99} in order to tackle the
verification problem of C11 programs under weak memory. We start by
developing the necessary language support for defining C11 programs
and have shown that existing operational semantics for the RC11-RAR
fragment~\cite{DBLP:conf/ppopp/DohertyDWD19} can be encoded in a
straightforward manner, which provides an example instantiation. We
have provided a set of proof rules to facilitate verification of C11
programs, and exemplified our approach by verifying a number of
example programs.

Our entire development has been carried out in the Isabelle theorem
prover and is modular with respect to the underlying memory model. For
the RC11-RAR fragment we consider, we have shown that the proofs are
highly automated. As described in
Section~\ref{subsect:verif-strategy}, a simple pattern of applying an
Owicki-Gries specific proof method, and then invoking SMT solvers via
Isabelle's sledgehammer tool was sufficient for verifying every proof
outline. Moreover, the use of Isabelle means that we have flexibility
in the specific operational semantics that we use.

\bibliographystyle{splncs04}
\bibliography{references}

 \appendix \newpage
\section{Appendix}
\label{sec:appendix1}

We present proofs of two additional variations of the RRC litmus
test. \reffig{fig:rrc2} presents a simple two-threaded version with a
writer thread that enforces a order of writes in program order and a
reader thread that reads from these writes. The proof shows that the
order of reads in the reader is consistent with the order of writes in
the writer.

\begin{figure}[h]
\begin{minipage}[b]{\columnwidth}
\small 
\begin{lstlisting}[escapeinside={*}{*},xleftmargin=2em, numbers=left, numberstyle=\tiny]
$\tt \color{blue} consts$ 
	x :: L
	
$\tt \color{blue} record$ mp_state =
	a :: V
	b :: V
	$\sigma$ :: Cstate

||- 
*\color{red}{$\ogal$ $\acute{}$a = 0 $\land$ $\acute{\;}$b = 0 $\ogar$}*
COBEGIN
*\color{red}{$\ogal$ [x =$_1$ 0] $\asigma$ $\land$ $[\zwr{\texttt{x}}{1}]$ $\asigma$ $\land$ [$\amo{\texttt{x}}{1}$] $\asigma$  $\land$ [$\zwr{\texttt{x}}{2}$] $\asigma$ $\land$  [$\amo{\texttt{x}}{1}$] $\asigma$ $\ogar$ }*
	<x :=$_1$ 1 $\asigma$>;;
*\color{red}{$\ogal$ [x =$_1$ 1] $\asigma$ $\land$ [$\enc{1}{\texttt{x}}{1}$] $\asigma$ $\land$ [$\zwr{\texttt{x}}{2}$]$_0$ $\asigma$ $\land$ [$\amo{\texttt{x}}{1}$] $\asigma$ $\land$ [$\amo{\texttt{x}}{2}$] $\asigma$ $\ogar$}*
	<x :=$_1$ 2 $\asigma$>
*\color{red}{$\ogal$ [$\amo{\texttt{x}}{1}$] $\asigma$ $\land$ [$\amo{\texttt{x}}{2}$] $\asigma$ $\land$ [1 $\dorder_\texttt{x}$ 2] $\asigma$$\ogar$}*
||
*\color{red}{$\ogal$  [$\amo{\texttt{x}}{1}$] $\asigma$ $\land$  [$\amo{\texttt{x}}{2}$] $\asigma$ $\ogar$}*
	< $\acute{}$a $\leftarrow$$_2$ x $\asigma$> ;;
*\color{red}{$\ogal$  [$\enc{2}{\texttt{x}}{\texttt{a}}$] $\asigma$ $\ogar$}*
	< $\acute{\;}$b $\leftarrow$$_2$ x $\asigma$> 
*\color{red}{$\ogal$  ($\acute{}$a $\neq$ $\acute{\;}$b $\longrightarrow$ [$\acute{}$a $\porder_\texttt{x}$ $\acute{\;}$b] $\asigma$) $\ogar$}*
COEND
*\color{red}{$\ogal$ $\acute{}$a = 2 $\longrightarrow$ $\acute{\;}$b $\neq$ 1 $\ogar$}*


\end{lstlisting}
 \caption[caption]{Isabelle encoding of the read-read coherence litmus test with two threads.}
 \label{fig:rrc2}
\end{minipage}
\end{figure}

\reffig{fig:rrc4} presents the standard four-threaded version in which
the two writes to {\tt x} occur in two different threads. There are
two reader threads both of which read from {\tt x} twice. One must
prove that the order of writes read by both threads are consistent. In
particular, if {\tt a} is set to 1 and {\tt b} to 2, then thread~3
must have seen the writes to {\tt x} in that order. It should
therefore not be possible for thread~3 to read 1 for {\tt x} if it has
already seen the value 2 in the first read at line 29.

\begin{figure}[t]

\begin{minipage}[b]{\columnwidth}
\small 
\begin{lstlisting}[escapeinside={*}{*}, xleftmargin=2em, numbers=left, numberstyle=\tiny]
$\tt \color{blue} consts$
	x :: L
	
$\tt \color{blue} record$ rrc4_state =
	a :: V
	b :: V
	c :: V
	d :: V
	$\sigma$ :: Cstate

||- 
*\color{red}{$\ogal$  [x $=_{1} 0$] $\asigma$ $\land$ [x $=_{2} 0$] $\asigma$ $\land$ [x $=_{3} 0$] $\asigma$ $\land$ [x $=_{4} 0$] $\asigma$ $\ogar$}*
COBEGIN
*\color{red}{$\ogal$  [$\zwr{\texttt{x}}{1}$]$_0$ $\asigma$ $\land$ [$\amo{\texttt{x}}{1}$] $\asigma$ $\ogar$}*
	*<x :=$_1$ 1  $\asigma$> *
*\color{red}{$\ogal$ True $\ogar$}*
||
*\color{red}{$\ogal$  [$\zwr{\texttt{x}}{2}$]$_0$ $\asigma$ $\land$ [$\amo{\texttt{x}}{2}$]  $\asigma$ $\ogar$}*
	*<x :=$_2$ 2  $\asigma$>*
*\color{red}{$\ogal$ True $\ogar$}*
||
*\color{red}{$\ogal$ True $\ogar$*
	*< $\acute{}$a $\leftarrow$$_3$ x  $\asigma$> ;;*
*\color{red}{$\ogal$  [$\enc{3}{\texttt{x}}{\acute{}\texttt{a}}$] $\asigma$ $\land$ [$\amo{\texttt{x}}{2}$] $\asigma$ $\land$ ($\acute{}$a  = 1 $\longrightarrow$ [$\amo{\texttt{x}}{\acute{}\texttt{a}}$] $\asigma$)  $\ogar$}*
	*< $\acute{\;}$b $\leftarrow$$_3$ x  $\asigma$> *
*\color{red}{$\ogal$($\acute{}$a $\neq$ $\acute{\;}$b $\longrightarrow$ [$\acute{}$a $\porder_\texttt{x}$ $\acute{\;}$b] $\asigma$) $\land$ ($\acute{}$a = 1 $\longrightarrow$ [$\amo{\texttt{x}}{\acute{}\texttt{a}}$] $\asigma$) $\land$ ($\acute{\;}$b = 2 $\longrightarrow$ [$\amo{\texttt{x}}{\acute{\;}\texttt{b}}$] $\asigma$)$\ogar$}*
||
*\color{red}{$\ogal$ True $\ogar$}*
	*< $\acute{}$c $\leftarrow$$_4$ x  $\asigma$> ;;*
*\color{red}{$\ogal$[$\enc{4}{\texttt{x}}{\acute{}\texttt{c}}$] $\asigma$ $\land$  ($\acute{}$c = 2 $\longrightarrow$ [$\amo{\texttt{x}}{\acute{}\texttt{c}}$] $\asigma$)$\ogar$}*
	*< $\acute{}$d $\leftarrow$$_4$ \texttt{x}  $\asigma$> *
*\color{red}{$\ogal$  ($\acute{}$c $\neq$ $\acute{}$d $\longrightarrow$ [$\acute{}$c $\porder_\texttt{x}$ $\acute{}$d] $\asigma$) $\land$ ($\acute{}$c = 2 $\longrightarrow$ [$\amo{\texttt{x}}{\acute{}\texttt{c}}$] $\asigma$) $\land$  ($\acute{}$d = 1 $\longrightarrow$ [$\amo{\texttt{x}}{\acute{}\texttt{d}}$] $\asigma$)$\ogar$}*
COEND
*\color{red}{$\ogal$ $\acute{}$a = 1 $\land$ $\acute{\;}$b = 2 $\land$ $\acute{}$c  = 2 $\longrightarrow$ $\acute{}$d $\neq$ 1 $\ogar$}*



\end{lstlisting}
 \caption[caption]{Isabelle encoding of the read-read coherence litmus test with four threads.}
 \label{fig:rrc4}
\end{minipage}

\end{figure}


\end{document}